\documentclass[showpacs,prd,nofootinbib,reprint,preprintnumbers]{revtex4-1}
\usepackage{adjustbox}
\usepackage[utf8]{inputenc}
\usepackage{color,graphicx,epsfig}
\usepackage{amsmath,amssymb,dsfont,epsfig,graphicx,xcolor,fontenc}
\usepackage{bm}
\usepackage[english]{babel}
\usepackage{graphicx}%
\usepackage{amsfonts}%
\usepackage{amssymb}
\usepackage{braket}
\usepackage{hyperref}
\usepackage{epstopdf}
\usepackage{ulem}
\usepackage{soul}
\usepackage{float}
\usepackage{multirow}
\usepackage[table]{xcolor}
\usepackage{slashed}
\usepackage{mathtools}
\usepackage{lmodern}
\usepackage{slantsc}
\usepackage{upgreek}

\definecolor{nicered}{rgb}{0.7,0.1,0.1}
\definecolor{nicegreen}{rgb}{0.1,0.5,0.1}
\definecolor{violet}{rgb}{0.7,0.3,0.3}
\hypersetup{colorlinks,citecolor= nicegreen,linkcolor= nicered}

\setstcolor{red}

\def\LjubljanaFMF{Faculty of Mathematics and Physics, University of Ljubljana,
 Jadranska 19, 1000 Ljubljana, Slovenia }
\def\LjubljanaIJS{Jo\v zef Stefan Institute, Jamova 39, 1000 Ljubljana, Slovenia}
\def\IFIC{Instituto de F\'isica Corpuscular,
Universitat de Val\'encia – Consejo Superior de Investigaciones Cient\'ificas,
Parc Cient\'ific, E-46980 Paterna, Valencia, Spain}

\begin{document}

\title{Dark light shining on $B\to K^{(*)} E_{\rm miss}$}

\author{Patrick~D.~Bolton}
\email[Electronic address: ]{patrick.bolton@ijs.si} 
\affiliation{\LjubljanaIJS}

\author{Jernej~F.~Kamenik}
\email[Electronic address: ]{jernej.kamenik@cern.ch} 
\affiliation{\LjubljanaIJS}
\affiliation{\LjubljanaFMF}

\author{Mart\'in~Novoa-Brunet}
\email[Electronic address: ]{martin.novoa@ific.uv.es}
\affiliation{\IFIC}

\begin{abstract}
Recent Belle~II data on $B^+ \to K^+ E_{\rm miss}$ show an excess consistent with a two-body decay involving a light invisible particle with mass around $2.1\,\mathrm{GeV}$. We present a UV-complete explanation based on a Higgsed $U(1)'$ gauge symmetry with a light vector boson $Z'$ and a vector-like top partner, which naturally enhances $b \to s$ transitions. While the minimal model can reproduce the required $B \to K^{(*)} Z'$ rate, it is excluded by LHCb searches for resonant dimuon decays due to unavoidable loop-induced couplings of $Z'$ to charged leptons. We show that a minimal extension with an additional light $U(1)'$-charged singlet fermion allows $Z'$ to decay dominantly invisibly, evades existing constraints coming also from dark photon and collider searches as well as Higgs measurements, and can simultaneously account for the Belle~II excess and the observed dark matter abundance through resonant thermal freeze-out.
\end{abstract}

\maketitle

%
\section{Introduction}
\label{sec:intro}
%

Rare flavour-changing neutral current (FCNC) transitions in the quark sector provide a powerful window into physics beyond the Standard Model (SM). The loop- and CKM-suppressed rates predicted in the SM allow in many cases to probe new short distance dynamics much above the TeV scale. However, new degrees of freedom affecting rare FCNC process can also appear at or even below the SM weak and QCD scales. The decays $B \to K^{(*)} E_{\rm miss}$, $D^0 \to \pi^0 E_{\rm miss}$ and $K \to \pi E_{\rm miss}$, where the missing energy represents final state particles that escape detection, have recently emerged as sensitive probes of such new physics. In particular, the Belle II collaboration has reported a $2.9\sigma$ excess in $B^+ \to K^+ E_{\rm miss}$ over SM expectations based on the $b \to s \nu \bar \nu$ transition~\cite{Belle-II:2023esi}. Even more interestingly, the excess is somewhat concentrated around the (two-body decay) kinematic region corresponding to the missing energy carried away by a single particle with mass of about $2.1\,\text{GeV}$~\cite{Bolton:2024egx}.\footnote{Very recently, {Belle} has presented preliminary results of their search targeting resonant $B^+\to K^+ (X \to \rm invisible)$ decays, putting stringent bounds on the two-body decay topology of $B^+ \to K^+ E_{\rm miss}$~\cite{deMarino2025_missingEnergyBelleII}. However, the region of resonant masses $m_X \sim 2$\,GeV preferred by our analysis~\cite{Bolton:2024egx} is vetoed due to overwhelming SM backgrounds and thus remains unconstrained.}

In Refs.~\cite{Bolton:2024egx,Bolton:2025fsq}, we analysed this excess using a model-independent approach based on the effective field theory (EFT) framework for new light SM singlet fields coupled to quark flavour-changing currents~\cite{Kamenik:2009kc} (see also Refs.~\cite{Fridell:2023ssf,Guedes:2024vuf,Buras:2024ewl,Rosauro-Alcaraz:2024mvx,Marzocca:2024hua,Hou:2024vyw} for related work). Among several considered scenarios which can consistently account for the Belle II result, the data give slight preference to the emission of a single spin-0 or spin-1 state, provided its mass and couplings to the quark sector fall within a narrow range. 
However, the EFT description alone leaves open the question of UV consistency and the microscopic origin of such scenarios. In this work, we try to address this by constructing an explicit, renormalisable new physics model, which matches onto the low energy EFT of a new massive vector field predominantly coupling to $b \to s$ currents. Our model is built on the aligned gauged and Higgsed $U(1)'$ framework with a vector-like top partner, following the general ideas introduced in Ref.~\cite{Kamenik:2017tnu}. 
It is an explicit realisation of a rank-1 flavour symmetry breaking~\cite{Gherardi:2019zil, Marzocca:2024hua} and naturally implements the minimally flavour violating approximate $U(2)$ flavour symmetry structure of BSM effects~\cite{DAmbrosio:2002vsn, Barbieri:2011ci, Barbieri:2012uh}, predicting largest contributions to $b \to s$ FCNCs.

Several alternative explanations of the Belle II excess within explicit models have appeared in the literature, targeting both two-body~\cite{McKeen:2023uzo,Altmannshofer:2023hkn,Berezhnoy:2023rxx,Altmannshofer:2024kxb,Ho:2024cwk,Calibbi:2025rpx,Berezhnoy:2025tiw, DiLuzio:2025qkc, Kim:2025zaf} and three-body~\cite{Datta:2023iln, Chen:2023wpb,He:2023bnk,Wang:2023trd, Felkl:2023ayn,Athron:2023hmz,Bhattacharya:2024clv, Buras:2024mnq, Becirevic:2024iyi,Hati:2024ppg, Kim:2024tsm,  He:2024iju,Chen:2024cll,Gabrielli:2024wys, Loparco:2024olo,Ho:2024cwk,Chen:2025npb, Crivellin:2025qsq} $B \to K^{} E_{\rm miss}$ decay kinematics. In particular, Ref.~\cite{Gabrielli:2024wys} considered a SM extension with an unbroken gauged $U(1)'$ symmetry and new light Dirac fermions charged under it. The Belle II excess is then accommodated via off-shell exchange of the massless dark photon associated with this new gauge symmetry, where the dark photon interactions to SM quarks are parametrised by non-renormalisable effective operators. Ref.~\cite{Ho:2024cwk} (see also Ref.~\cite{Kim:2025zaf}) instead considers a similar model, where the extra symmetry is a (Higgsed) $U(1)_{\tau - \mu}$. In this case the Belle II signal is accommodated via light $U(1)_{\tau - \mu}$ Higgs boson emission (either on- or off-shell) and subsequent decay to new fermions, also charged under $U(1)_{\tau - \mu}$. The corresponding massive vector boson in this model does not play a relevant role in flavour phenomenology but is crucial in accounting for correct cosmological dark matter (DM) abundance of the new fermions. Similarly, Ref.~\cite{Altmannshofer:2023hkn} briefly considers a Higgsed $U(1)_{B_3 - L_3}$, where the emission of the associated light $Z'$ decaying to $\nu_\tau \bar \nu_\tau$ pairs accounts for the Belle II excess. Conversely, Ref.~\cite{DiLuzio:2025qkc} considered gauging and Higgsing the anomalous $U(1)_\tau$, where the $Z'$ couplings to SM quarks arise at one-loop through the associated Wess-Zumino terms~\cite{DiLuzio:2025qkc}.
Finally, Ref.~\cite{Calibbi:2025rpx} recently considered a Higgsed $U(1)'$ mediating DM interactions, where the associated light $Z'$ appearing in $B \to K^{(*)} E_{\rm miss}$ only interacts with the SM fermions though kinetic mixing with hypercharge.

The main goals of this paper are threefold. First, we present the minimal aligned $U(1)'$ model realisation and demonstrate how it could account for the Belle II excess in $B \to K^{(*)} E_{\rm miss}$ through the production of a light massive and invisibly decaying $Z'$. We also comment on similarities and main differences with existing related proposals~\cite{Altmannshofer:2023hkn, DiLuzio:2025qkc, Calibbi:2025rpx}. Second, we analyse the full set of phenomenological constraints on the model, including other flavour observables, electroweak precision data and collider bounds.
Importantly, we show that after electroweak symmetry breaking, irreducible $Z - Z' $ mass-mixing induces $Z'$ couplings to all SM fermions and the model is excluded by the LHCb searches for prompt and displaced $\mu^+\mu^-$ resonances in $B \to K^{(*)} \mu^+ \mu^-$ decays~\cite{LHCb:2015nkv,LHCb:2016awg}.
Third, we investigate a possible minimal extension of the model that avoids the LHCb constraint by introducing an additional light SM singlet fermion charged under the $U(1)'$ and thus opening new dominant invisible decay channels of the $Z'$. This additional particle turns out to be a viable DM candidate and we explore the related phenomenology in detail.

Altogether, our results highlight the potential of UV-complete dark sector models to address the Belle II anomaly while remaining predictive and testable across multiple experimental frontiers.

The remainder of this work is organised as follows. In Sec.~\ref{sec:minimal}, we introduce the minimal aligned $U(1)'$ model, detailing its field content, gauge and scalar interaction sectors, and the resulting mass and mixing structures. We then study its implications for $B \to K^{(*)} E_{\rm miss}$ at Belle II, together with other closely related experimental constraints. In Sec.~\ref{sec:minimal_DM}, we study a minimal extension of the model with an additional light SM singlet fermionic field introducing interesting correlations between flavour probes, dark photon and collider searches, electroweak precision observables, and DM phenomenology. 
We summarise our findings and discuss possible other, non-minimal extensions of the model in Sec.~\ref{sec:conclusions}.

%
\section{Minimal Model}
\label{sec:minimal}
%

Here, we consider a minimal aligned $U(1)'$ model which introduces a vector-like quark $T'({\bf 3} ,{\bf 1},2/3,q')$ and a SM singlet scalar $\Phi({\bf 1},{\bf 1},0,q')$, where $q'$ is the charge of both fields under an additional $U(1)'$. The Lagrangian of the model contains
\begin{align}
\label{eq:model}
\mathcal{L} &\supset - \frac{1}{4} B'_{\mu\nu} B'^{\mu\nu} - \frac{\epsilon_B}{2} B_{\mu\nu} B'^{\mu\nu} + \big|D_\mu \Phi\big|^2 - V(H, \Phi) \nonumber \\
&\hspace{1.5em}  + \bar{T}'(i\slashed{D} - M_T)T' - \Big[y_T^r \bar{T'} \Phi u_{Rr}' + \text{h.c.}\Big] \,,
\end{align}
with $D_\mu \supset i \tilde{g} q' B_\mu'$ and $B_{\mu\nu}' = \partial_\mu B_\nu' - \partial_\nu B_\mu'$, where $B_\mu'$ is the gauge field associated with the $U(1)'$, $B_\mu$ is the $U(1)_Y$ hypercharge gauge field, $H$ the $SU(2)_L$ Higgs doublet, and $u'_{Rr}$ the up-type right-handed quark with generation index $r=u,c,t$. Additional terms in the scalar potential are
\begin{align}
\label{eq:potential}
V(H,\Phi) = -\tilde{\mu}^2 |\Phi|^2 + \tilde{\lambda} |\Phi|^4 + \lambda' |\Phi|^2 (H^\dagger H)\,.
\end{align}
The complex scalar $\Phi$ obtains a vacuum expectation value (vev) $\tilde{v}$ that breaks the $U(1)'$. In the broken phase, the model admits an additional massive vector boson, $Z'$, with the mass $M_{Z'} = \tilde{g}q'\tilde{v}$.

%
\subsection{Spectrum and Theoretical Constraints}
\label{subsec:model}
%

\subsubsection{Fermions}

In the following, we consider $y_T^t \neq 0$ and $y_T^u = y_T^c = 0$, so that the vector-like quark $T'$ only mixes with the top quark (and is thus a heavy top partner). In the broken phase, the relevant mass matrix in the top-type quark sector is thus
\begin{align}
\label{eq:mass_matrix}
\mathcal{L} \supset 
-\frac{1}{\sqrt{2}}\begin{pmatrix}
\bar{t}_{L}' & \bar{T}_L'
\end{pmatrix}
\begin{pmatrix}
vy_t & 0 \\
\tilde{v}y_T^t & \sqrt{2}M_T \\
\end{pmatrix}
\begin{pmatrix}
t_{R}' \\ T_R'
\end{pmatrix} + \text{h.c.} \,,
\end{align}
with $y_t$ the top Yukawa coupling, $\mathcal{L}\supset - y_t \bar{Q}_3\tilde{H}t_R' + \text{h.c.}$, and where, without loss of generality, $y_t$, $y_T^t$ and $M_T$ can be made real and positive by a rephasing of the chiral fields. The resulting mass basis fields can be obtained via the bi-unitary rotation $t_{X}' = P_{X}(c_X t + s_X T)$ and $T_{X}' = P_{X}(-s_X t + c_X T)$ for $X = L,R$, with $s_X = \sin\theta_X$ and $c_X = \cos\theta_X$, where the left- and right-handed mixing angles are given by
\begin{align}
\label{eq:mixing}
\tan2\theta_{L,R} &= \frac{\tilde{v} y_T^t\big\{v y_t, \sqrt{2}M_T\big\}}{M_T^2 \pm (\tilde{v} y_T^t)^2/2 -(v y_t)^2/2} \,,
\end{align}
and the physical masses satisfy $m_t m_T = vy_t M_T/\sqrt{2}$ and $m_t^2 + m_T^2 = (vy_t)^2/2 + (\tilde{v}y_T^t)^2/2 + M_T^2$. In the following, $\{m_T,\tilde{g}q', y_T^t\}$ are taken as input parameters, assuming $m_T > m_t$. Using the relation $M_{Z'} = \tilde{g}q'\tilde{v}$, the vev $\tilde{v}$ can be eliminated in favour of $\tilde{g}q'$ and $M_{Z'} = 2.1$~GeV, with the latter being kept fixed to best fit of the Belle~II excess. The top quark Yukawa coupling is given in terms of the input parameters as $vy_t/\sqrt{2} = m_t c_L c_R + m_T s_L s_R$. From the identity $m_t c_L s_R = m_T c_R s_L$, we expect $s_L \ll s_R$ in the limit $m_t \ll m_T$. For $y_T^t \to 0$, $s_{L,R}\to 0$ and $t$ (with $m_t = vy_t/\sqrt{2}$) and $T$ ($m_T = M_T$) decouple.

In most of the viable parameter space, the $t-T$ mixing is given by $s_R \approx s_L m_T /m_t \approx \tilde{v} y_T^t/\sqrt{2}m_T$. However, we can determine the maximal values of $s_{L,R}$ by combining the relations below Eq.~\eqref{eq:mixing} with $(M_T - v y_t/\sqrt{2})^2 \geq 0$, yielding $\tilde{v}y_T^t/\sqrt{2}\leq m_T - m_t$, or equivalently
\begin{align}
\label{eq:consistency}
\tilde{g}q' \geq 
\frac{M_{Z'} y_T^t}{\sqrt{2}(m_T - m_t)} \,.
\end{align}
We refer to this as a \textit{consistency} bound, as its violation would imply deviations from the assumed $t-T$ mixing from Eq.~\eqref{eq:mass_matrix}. The lower bound in Eq.~\eqref{eq:consistency} corresponds to the upper limits $s_{L,R}^2 \leq m_{t,T}/(m_t + m_T)$. To remain compatible with perturbative unitarity, the two Yukawa couplings must satisfy $y_t, y_T^t < \sqrt{8\pi/3}$~\cite{Allwicher:2021rtd}.

Considering next the quark gauge interactions in the broken phase, i.e. $\mathcal{L} \supset 
-\mathcal{J}_\mu \mathcal{A}^\mu -\frac{g}{\sqrt{2}}[J_\mu^+ W^{+\mu} +\text{h.c.}]$,
where $\mathcal{J}_\mu = (eJ_\mu,~\frac{g}{c_w}J_\mu^Z,~\tilde{g}J_\mu')$ and $\mathcal{A}_\mu = (A_\mu,~Z_\mu,~Z_\mu')^T$, the up-type quarks can be rotated to the mass basis. The currents then contain the terms
\begin{align}
\label{eq:top_currents}
J_\mu &\supset Q_u (\bar{t}\gamma_\mu t + \bar{T}\gamma_\mu T) \,, \nonumber \\
J_\mu^Z &\supset \frac{1}{2} (c_L\bar{t} + s_L \bar{T})\gamma_\mu P_L (c_L t + s_L T) - s_w^2 J_\mu \,, \nonumber\\
J_\mu' & \supset q'(- s_X \bar{t} + c_X \bar{T}) \gamma_\mu P_X (-s_X t + c_X T) \,, \nonumber \\
\hspace{-0.5em}J_\mu^+ &\supset V_{ti}(c_L\bar{t} + s_L \bar{T}) \gamma_\mu P_L d_{i} \,,
\end{align}
where $Q_u = 2/3$, $\theta_w$ is the weak mixing angle, and $V_{ij}$ are elements of the Cabibbo–Kobayashi–Maskawa (CKM) matrix.

\subsubsection{Gauge Bosons}
\label{subsubsec:gauge}

\begin{figure}[t!]
  \centering
  \includegraphics[width=0.435\columnwidth]{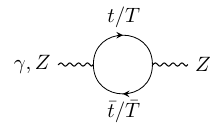}
  \caption{One-loop contributions of $t/T$ to the kinetic and mass mixing between $Z'$ and $\gamma/Z$.}
  \label{fig:kinetic-mixing}
\end{figure}

In general, $A_\mu$, $Z_\mu$ and $Z'_\mu$ above are not gauge sector mass eigenstates of the model. At tree-level, the SM gauge bosons $B/W^{3}$ do not mix with $B'$. The former are rotated to the mass basis fields $\gamma$/$Z$, while $B'$ produces the extra massive field $Z'$. However, at one-loop, kinetic and mass mixing of $Z'$ and $\gamma/Z$ arises. The kinetic mixing of $B'$ and $B$ is controlled by $\epsilon_B$ in Eq.~\eqref{eq:model}, which runs above the vector-like quark mass $M_T$ according to the renormalisation group evolution (RGE) $d\epsilon_B/d\ln\mu = - e \tilde{g}q'/3\pi^2c_w$. In the unbroken phase, this running is generated by the one-loop exchange of $T'$, which is charged under both $U(1)'$ and $U(1)_Y$. The RGE also applies in the broken phase of the theory, with the mass eigenstates $t/T$ in the loop, as depicted in Fig.~\ref{fig:kinetic-mixing}. We note that in the broken phase, the kinetic mixing $\epsilon_W$ is also effectively induced between $B'$ and $W^3$ from the mixing of upper component of the $SU(2)_L$ doublet $Q_3$ and $T_L'$. This is is equivalent to the contribution of the $d = 8$ operator $|\Phi|^2(H^\dagger \tau^3 H)W_{\mu\nu}^3 B'^{\mu\nu}$~\cite{Bauer:2022nwt}. The kinetic mixing $\epsilon_B$ also receives threshold contributions encoded by $|\Phi|^2 B_{\mu\nu} B'^{\mu\nu}$  at $d = 6$ and $|\Phi|^2 (H^\dagger H) B_{\mu\nu} B'^{\mu\nu}$ and $|\Phi|^4 B_{\mu\nu} B'^{\mu\nu}$ at $d = 8$.

For $\epsilon_{B,W}\neq 0$, the kinetic mixing terms in the gauge boson mass basis are $\mathcal{L} \supset -\frac{\epsilon_A}{2} F_{\mu\nu}Z'^{\mu\nu} - \frac{\epsilon_Z}{2} Z_{\mu\nu}Z'^{\mu\nu}$, where $\epsilon_A = c_w \epsilon_B + s_w \epsilon_W$ and $\epsilon_Z = - s_w \epsilon_B + c_w \epsilon_W$ are the $\theta_w$ rotated kinetic mixing coefficients. For $\epsilon_B(\mu) \neq 0$ at the high scale $\mu =  \Lambda \gg m_T$, $\epsilon_{A,Z}$ are given by
\begin{align}
\label{eq:eps_A}
\epsilon_A &= c_w\epsilon_B(\Lambda) + \frac{e\tilde{g}q'}{6\pi^2}\bigg[\ln\frac{\Lambda^2}{m_T^2}+\frac{1}{2}(s_L^2 + s_R^2)\ln r\bigg] \,, \\
\epsilon_Z & = - t_w \epsilon_A - \frac{g\tilde{g}q'}{16\pi^2 c_w} c_L^2 s_L^2 f(r) \,,
\label{eq:eps_Z}
\end{align}
where $t_w = \tan\theta_w$, $r = m_T^2/m_t^2$, and
\begin{align}
f(x) &= \frac{5x^2 - 22x + 5}{3(x - 1)^2} - \frac{x^3 - 3x^2 - 3x + 1}{(x - 1)^3}\ln x \,.
\label{eq:f_loop_func}
\end{align}
The second term in $\epsilon_Z$ arises from threshold effects. In the following, we will mostly assume that $\epsilon_B(\Lambda) = 0$ at the high scale $\Lambda = 10^{16}$~GeV and use the induced values of $\epsilon_{A,Z}$ from the RGE plus threshold corrections. In much of the relevant parameter space, $s_{L} \ll s_R < 1$, giving
$\epsilon_A \approx -\epsilon_Z/t_w \approx 3\times 10^{-4}(\tilde{g} q'/10^{-3})(L_T/55)$ with $L_T \equiv \ln \Lambda^2/m_T^2$. We note that $\epsilon_A$ may be suppressed if there is a cancellation between $\epsilon_B(\Lambda)$ and the value generated by the running. Even in the presence of this fine-tuning, $\epsilon_Z$ would still be generated by the second term in Eq.~\eqref{eq:eps_Z}. Finally, in the $y_T^t\to 0$ limit, $\epsilon_A = -\epsilon_Z/t_w = e\tilde{g}q'L_T/6\pi^2$ holds exactly.

After electroweak symmetry breaking, the diagram in Fig.~\ref{fig:kinetic-mixing} also induces mass mixing between $Z$ and $Z'$. The effective mass term can be written as $\mathcal{L} \supset M_{ZZ'}^2 Z_\mu Z'^{\mu}$, from which it is convenient to define the dimensionless quantity $\delta_Z = M_{ZZ'}^2/M_{Z}^2$. For $m_t \ll m_T$, the one-loop contribution yields
\begin{align}
\label{eq:delta_Z}
\delta_Z = -\frac{3g\tilde{g}q'}{32\pi^2c_w}\frac{m_t^2}{M_Z^2} s_L^2 \Big[c_L^2 g(r) - 2 c_R^2 h(r)\Big]\,,
\end{align}
where $g(x) = 1 + x - \frac{2x}{x-1}\ln x$ and $h(x) = 2x - \frac{x(x + 1)}{x-1}\ln x$. The mass mixing can also be viewed as the contribution of the $d = 6$ operator $(H^\dagger D_\mu H)(\Phi^\dagger D^\mu \Phi) + \text{h.c.}$, and is an irreducible effect (cannot be cancelled against a UV parameter) in the model for $y_T^t\neq 0$. Thus $\delta_Z \to 0$ is only possible for $y_T^t \to 0$, while on the other hand we need $y_T^t\neq 0$ in order to accommodate the Belle~II result via $B^+\to K^+ Z'$ decays, as we demonstrate in Sec.~\ref{subsec:belleii}.

The impact of the kinetic and mass mixing can be seen in the low energy interactions of $\gamma$, $Z$ and $Z'$. With $\epsilon_{A,Z}\neq 0$, the massive gauge fields are not canonically normalised; a non-unitary rotation can first diagonalise the matrix containing the kinetic mixing terms, followed by a unitary rotation diagonalising the gauge boson mass matrix containing the mass mixing term $\delta_Z$. We obtain the following redefinition of fields: $A_\mu \to A_\mu -\epsilon_A Z'_\mu$, $Z_\mu \to Z_\mu - (\delta_Z - x \epsilon_Z) Z'_\mu$ and $Z'_\mu \to Z'_\mu + (\delta_Z - \epsilon_Z) Z_\mu$,
where we expand to leading order\footnote{For the redefinitions of $A_\mu$ and $Z'_\mu$, the leading order terms are approximately accurate up to $\tilde{g}q'\lesssim 1$. However, for $Z_\mu$, higher order terms in the expansion, e.g. $\mathcal{O}(\epsilon_Z^3)$, become relevant for $\tilde{g}q' \gtrsim 10^{-2}$. On the other hand, the implied values of $\epsilon_A$ for $\tilde{g}q' \gtrsim 10^{-2}$ are excluded by searches for dark photons, as discussed in Sec.~\ref{subsubsec:darkPhoton}. Thus, we will only consider $\tilde{g}q' < 10^{-2}$ in this work, so that the leading order terms in the field redefinitions remain dominant.} in $\epsilon_{A,Z}$, $\delta_Z$ and the ratio $x \equiv M_{Z'}^2/M_Z^2 = 5.3\times 10^{-4}$. Explicitly, the coupling of $Z'$ to fermions is $\mathcal{L} \supset - \bar{f} \gamma_\mu \big(v_{Z'}^f - a_{Z'}^f \gamma_5\big) f Z'^{\mu}$, with the vector and axial-vector couplings
\begin{align}
v_{Z'}^f &= \tilde{g}\mathfrak{v}^f - e Q_f \epsilon_A   - \frac{g}{2c_w} (T_3^f \vartheta^f - 2s_w^2 Q_f) \Delta_{Z'} \,, \nonumber \\
a_{Z'}^f &= \tilde{g}\mathfrak{a}^f - \frac{g}{2c_w} T_3^f \vartheta^f \Delta_{Z'}\,,
\label{eq:Zp_vector_axialvector}
\end{align}
where $T_3^f$ is the third component of $SU(2)_L$ isospin, $Q_f$ is the $U(1)_{\text{em}}$ charge, and we define $\Delta_{Z'} = \delta_Z-x\epsilon_Z$. The couplings in Eq.~\eqref{eq:Zp_vector_axialvector} are diagonal in flavour space for $f = \nu, \ell, d, u$ ($i,j \neq t,T$), with $\mathfrak{v}_{ij}^f, \mathfrak{a}_{ij}^f = 0$ and $\vartheta_{ij}^f = \delta_{ij}$. For up-type quarks with $i,j = t,T$, the couplings can be read off from Eq.~\eqref{eq:top_currents}, i.e.
\begin{align}
\{\mathfrak{v}_{tt}^u,\mathfrak{a}_{tt}^u\} &= q'(s_L^2 \pm s_R^2)/2 \,, \nonumber \\
\{\mathfrak{v}_{TT}^u,\mathfrak{a}_{TT}^u\} &= q'(c_L^2 \pm c_R^2)/2 \,, \nonumber \\
\{\mathfrak{v}_{tT}^u,\mathfrak{a}_{tT}^u\} &= \{\mathfrak{v}_{Tt}^u,\mathfrak{a}_{Tt}^u\} = - q'(s_L c_L \pm s_R c_R)/2 \,,
\end{align}
and $\vartheta_{tt}^u = c_L^2$, $\vartheta_{TT}^u = s_L^2$, and $\vartheta_{tT}^u = \vartheta_{Tt}^u = c_L s_L$. 
For $\epsilon_{A,Z} \gg \delta_Z$ and $x\ll 1$, the couplings of SM fermions to $Z'$ are effectively proportional to $e Q_f$, corresponding to the dark photon limit. Instead for $\epsilon_{A,Z} \ll \delta_Z$, $Z'$ possesses SM $Z$-like couplings. Consequently, the kinetic and mass mixing determine the $Z'$ lifetime, as examined in Sec.~\ref{subsec:belleii}. Finally, the modified $Z$ couplings to fermions read $\mathcal{L} \supset - \bar{f} \gamma_\mu \big(v_{Z}^f - a_{Z}^f \gamma_5\big) f Z^{\mu}$, with the vector and axial-vector couplings
\begin{align}
v_{Z}^f &= \frac{g}{2c_w}(T_3^f \vartheta^f - 2s_w^2 Q_f) + \tilde{g} \mathfrak{v}^f \Delta_Z \,, \nonumber\\
a_{Z}^f &= \frac{g}{2c_w}T_3^f \vartheta^f + \tilde{g} \mathfrak{a}^f \Delta_Z \,,
\label{eq:Z_vector_axialvector}
\end{align}
where the $\mathfrak{v}^f$, $\mathfrak{a}^f$ and $\vartheta^f$ factors are the same as above, and we define $\Delta_Z = \delta_Z - \epsilon_Z$. 

\subsubsection{Scalars}

Next we consider the scalar boson sector of the minimal model. The full scalar potential is minimised for $\mu^ 2 =  \lambda v^2 + \lambda' \tilde{v}^2/2$ and $\tilde{\mu}^2 = \tilde{\lambda} \tilde{v}^2 + \lambda' v^2/2$, where $\mu^2$ and $\lambda$ are the usual Higgs mass parameter and quartic coupling, $V\supset - \mu^2 H^\dagger H + \lambda (H^\dagger H)^2$, and $v$ is the Higgs vev. For the scalar potential to be bounded from below, the quartic couplings must satisfy $\lambda\,, \tilde{\lambda}\,, 4\lambda \tilde{\lambda} - \lambda'^2 \geq 0$. Expanding the scalar fields around the scalar potential minimum as $H = (\chi^+~(v+h'+i\chi')/\sqrt{2})^T$ and $\Phi = (\tilde{v} + \phi' + i \eta')/\sqrt{2}$, the extended CP-even scalar mass matrix in the broken phase is
\begin{align}
\label{eq:scalar_sector}
\mathcal{L} &\supset - \frac{1}{2}
\begin{pmatrix}
h' & \phi' 
\end{pmatrix}
\begin{pmatrix}
2\lambda v^2 & \lambda'v \tilde{v} \\
\lambda'v \tilde{v} & 2 \tilde{\lambda}\tilde{v}^2
\end{pmatrix}
\begin{pmatrix}
h' \\
\phi'
\end{pmatrix}\,.
\end{align}
We diagonalise it via the unitary rotation $h' = c_\phi h + s_\phi \phi$ and $\phi' = -s_\phi h + c_\phi \phi$, with $\tan 2\theta_\phi = \lambda' v \tilde{v}/(\tilde{\lambda}\tilde{v}^2 - \lambda v^2)$.
The physical scalar masses are then related to the quartic couplings and vevs as $(M_h M_{\phi})^2 = (4 \lambda\tilde{\lambda} - \lambda'^2)(v \tilde{v})^2$ and $M_h^2 + M_{\phi}^2 = 2(\lambda v^2 + \tilde{\lambda} \tilde{v}^2)$, and therefore $\lambda$ and $\tilde{\lambda}$ can be expressed in terms of $M_h = 125.2~\text{GeV}$, $v = 246~\text{GeV}$ and inputs $\{M_\phi, \tilde{g}q', \lambda'\}$. The inequality $2 \lambda' v \tilde{v}  \leq |M_\phi^2 - M_h^2|$ is furthermore implied. 

At one-loop order, $\lambda'$ receives the RGE contributions $d \lambda'/d\ln\mu = - 3 (y_t y_T^{t})^2/4\pi^2$, where we retain only the dominant box diagram expression. Fixing $\lambda'(\Lambda)=0$ in the UV, the quartic coupling at scales below $m_T$ can then be estimated, including threshold contributions, for $m_t \ll m_T$ as
\begin{align}
\label{eq:lambda_prime}
\lambda' &= \frac{3(y_t y_T^{t})^2}{8\pi^2}\bigg[\ln \frac{\Lambda^2}{m_T^2} + 1\bigg] \,,
\end{align}
which yields $\lambda' = 2 \times 10^{-2}(y_T^{t}/0.1)^2(L_T/55)$, assuming $y_t \approx \sqrt{2}m_t/v$. In this work, we will take $\lambda'(\Lambda) = 0$ at the scale $\Lambda = 10^{16}$~GeV and use the value of $\lambda'$ generated by the running to low scales, rather than treating it as a free parameter. Consequently, in much of the relevant parameter space, it is not possible to satisfy the condition $2\lambda'v\tilde{v} \leq |M_\phi^2 - M_h^2|$ in the limit $M_\phi \ll M_h$. For simplicity, in the following we will therefore only consider the $M_\phi > M_h$ scenario.\footnote{Note that in the minimal model, $Z'\to \phi \phi$ decays, even if kinematically allowed, are forbidden by parity, and thus the presence of a light $\phi$ with $M_\phi\lesssim 1$\,GeV would not change the main conclusions presented in Sec.~\ref{subsec:belleii}.} 

Another constraint on the scalar sector of the model arises from perturbative unitarity. The amplitudes for scattering processes $a \to b$ involving CP-even initial and final states $a, b \in \{W^+W^-, ZZ, h'h', \phi'\phi',Z'Z'\}$, can be decomposed into partial waves to build a coupled-channel matrix of $J = 0$ s-wave amplitudes,
\begin{align}
\label{eq:coupled_channel_matrix}
a_0=-\frac{1}{16\pi}
\begin{psmallmatrix}
4\lambda & \sqrt{2}\lambda & \sqrt{2}\lambda & 0 & 0 \\
\sqrt{2}\lambda & 3\lambda & \lambda & 0 & 0
\\
\sqrt{2}\lambda & \lambda & 3\lambda & \lambda'/2 & 0 \\
0 & 0 & \lambda'/2 & 3\tilde{\lambda} & \tilde{\lambda} \\
0 & 0 & 0 & \tilde{\lambda} & 3\tilde{\lambda}
\end{psmallmatrix}\,,
\end{align}
where initial/final state masses have been neglected and the limit of large $s$ is taken. The condition $|\text{Re} \, a_0| \leq 1/2$ for perturbative unitarity~\cite{Lee:1977eg,Lee:1977yc} is then applied to the largest eigenvalue of Eq.~\eqref{eq:coupled_channel_matrix}.
If $\lambda'$ is sufficiently small, the SM and $U(1)'$ sectors decouple and the subsequent bounds are $\lambda < 4\pi/3$ and $\tilde{\lambda} < 2\pi$. These upper bounds can be combined with the relations below Eq.~\eqref{eq:scalar_sector} to derive an approximately allowed range of $\phi$ masses,
\begin{align}
\label{eq:Mphi_allowed}
\frac{8\pi v^2}{3} + \frac{3(\lambda'\tilde{v})^2}{8\pi A} < M_\phi^2 < 4\pi\tilde{v}^2 + \frac{(\lambda' v)^2}{4\pi B}\,,
\end{align}
where $A = 1 - 3M_h^2/8\pi v^2$, $B = 1 - M_h^2/4\pi\tilde{v}^2$. The lower bound on $M_\phi$ is valid only for $\lambda' > 8\pi vA /3 \tilde{v}$; otherwise, $M_\phi$ is compatible with perturbative unitarity down to the lower limit $M_\phi^2 \geq M_h^2 + 2\lambda'v\tilde{v}$. When the lower and upper limits coincide, no value of $M_\phi$ can satisfy the condition $|\text{Re} \, a_0| \leq 1/2$. Solving this limit for $\lambda'$ then indicates that $\lambda' < 4\pi \,\text{min}[\tilde{v}B/v,\sqrt{2AB/3}]\approx 4\pi \,\text{min}[\tilde{v}/v,\sqrt{2/3}]$ is required for perturbative unitarity in the scalar sector, where we neglected $M_h$ in the second step. Finally, the lower bound $\tilde{v} > M_h/\sqrt{4\pi}\approx 35~\text{GeV}$ is always satisfied for $M_\phi > M_h$, implying $\tilde{g}q' < 0.06$ for $M_{Z'} = 2.1~\text{GeV}$.

\subsection{Phenomenology in $B$ Decays}
\label{subsec:belleii}

%
\begin{figure}[t!]
  \centering
  \includegraphics[width=0.65\columnwidth]{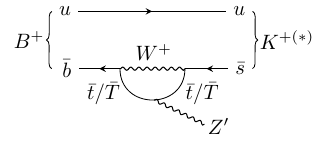}
  \caption{Two-body decay contributing to $B^+ \to K^{+}E_{\text{miss}}$, with a single light $Z'$ in the final state.}
  \label{fig:BtoKZprime}
\end{figure}

At scales relevant for $B$ decays, couplings of $Z'$ to down-type quarks are generated by one-loop diagrams such as in Fig.~\ref{fig:BtoKZprime}. The relevant effective Hamiltonian in the weak effective theory (WET) plus $Z'$ is of the form
\begin{align}
\label{eq:eff_Hamiltonian}
\mathcal{H}_{\text{eff}} &\supset C_{ij}^{V} \mathcal{O}_{ij}^{V} + C_{ij}^{\prime V} \mathcal{O}_{ij}^{\prime V} + \Big[C_{ij}^{T} \mathcal{O}_{ij}^{T} + \text{h.c.}\Big] \,,
\end{align}
where the vector and dipole-like operators are
\begin{align}
\label{eq:vector_coupling}
\mathcal{O}_{ij}^{(\prime)V} &= (\bar{d}_i \gamma_\mu P_{L(R)} d_j) Z'^\mu \,,\\
\mathcal{O}_{ij}^{T} &= (\bar{d}_i \sigma_{\mu\nu} P_L d_j) Z'^{\mu\nu} \,,
\label{eq:tensor_coupling}
\end{align}
respectively. The coefficients $C(\mu)$ can be matched to the full model at the electroweak scale, $\mu \sim M_W$, and then run down to $\mu \sim m_b$ under $SU(3)_c\times U(1)_\text{em}$. The complete expressions for the coefficients $C(M_W)$ which we use in our numerical studies are provided in App.~\ref{app:exps}.

It is instructive to understand important properties and limits of $C(\mu)$. First, the right-handed vector coupling vanishes identically, $C_{ij}^{\prime V}(M_W) = 0$. For $m_t \ll m_T$, the left-handed vector coupling at $\mu = M_W$ can be approximated as
\begin{align}
\label{eq:CV}
C_{ij}^{V}(M_W) &\approx \frac{1}{2} \mathcal{C}_{ij} m_t^2\bigg[\tilde{g}q' s_R^2 X_t - \frac{g}{c_w} \Delta_{Z'} Y_t\bigg]\,,
\end{align}
with $X_t = \ln m_T^2/M_W^2 -1$, $Y_t \equiv Y(m_t^2/M_W^2) = 1.47$ using the function $Y(x)$ defined in App.~\ref{app:exps} and we define for convenience $\mathcal{C}_{ij} = 4G_FV_{ti}^*V_{tj}/16\sqrt{2}\pi^2$. The first term in Eq.~\eqref{eq:CV} arises from the direct coupling of $Z'$ to $t/T$ in Eq.~\eqref{eq:Zp_vector_axialvector}, while the second term is induced by the mixing of $Z'$ into $Z$. The mixing of $Z'$ into $\gamma$ does not contribute, because the photon coupling to the FCNC down-type quark vector current vanishes due to gauge invariance. For $i = j$, the $Z'$ also couples to down-type quarks according to Eq.~\eqref{eq:Zp_vector_axialvector}. The expression in Eq.~\eqref{eq:CV} only holds for $\tilde{g}q'$ sufficiently above the consistency bound in Eq.~\eqref{eq:consistency}. To fit the Belle~II excess in some of the parameter space, we find that $\tilde{g}q'$ is required to be close to the consistency bound. In this limit, the left-handed vector coupling is instead given by
\begin{align}
\label{eq:CV_limit}
C_{ij}^{V}(M_W) &\approx \mathcal{C}_{ij}m_t m_T\tilde{g}q'\,.
\end{align}
The vector operators do not run under QED or QCD, so the expressions for $C_{ij}^{V}$ and $C_{ij}^{\prime V}$ above and in App.~\ref{app:exps} are also valid at $\mu \sim m_b$.

In the limit $m_t\ll m_T$, the dipole-like operator is given approximately at the matching scale $\mu = M_W$ by
\begin{align}
\label{eq:CT}
\hspace{-0.2em}C_{ij}^{T}(M_W) &\approx m_i \mathcal{C}_{ij}\Big[\tilde{g}q's_R^2 A_t - e\epsilon_A B_t - \frac{g}{c_w}\Delta_{Z'} C_t\Big]\,,
\end{align}
where $A_t = 0.0323$, $B_t = 0.194$ and $C_t = 0.150$ are the numerical outputs of loop functions, given in App.~\ref{app:exps}. Like the vector coupling $C_{ij}^V$, the first term in Eq.~\eqref{eq:CT} is induced by $Z'$ coupling directly to $t/T$. The second and third terms in Eq.~\eqref{eq:CT} are generated by the mixing of $Z'$ into $\gamma$ and $Z$, respectively. The running of $C_{ij}^{T}$ down to $\mu\sim m_b$ is dominated by QCD corrections. However, a detailed treatment of the running is not necessary for the following analysis, as we will demonstrate shortly.

Through these vector and dipole-like interactions, the $Z'$ can be produced on-shell in the process $B^+\to K^+Z'$ for $M_{Z'} < m_B - m_K$, with the branching fraction
\begin{align}
\label{eq:BtoKZp}
\mathcal{B}(B^+ \to K^+ Z') &= \tau_{B^+}\frac{|\vec{p}_K|^3}{8\pi M_{Z'}^2}\Big|C_{bs}^{V} f_+ + C_{bs}^T \tilde{f}_T \Big|^2\,,
\end{align}
where $\tau_{B^+}$ is the $B^+$ lifetime, $\vec{p}_K$ is the three-momentum of $K^+$ in the $B^+$ rest-frame, and $\tilde{f}_T = 2M_{Z'}^2 f_T/(m_B + m_K)$. Here, $f_+$ and $f_T$ are the vector and tensor $B\to K$ transition form factors evaluated at the squared momentum transfer $q^2 = M_{Z'}^2$, respectively. We note that the coupling $C_{sb}^{T}$ also contributes to the process, but is suppressed by $m_s/m_b$ with respect to $C_{bs}^{T}$ and is therefore neglected. The $Z'$ can also be produced on-shell in the process $B \to K^{*}Z'$ for $M_{Z'} < m_B - m_{K^*}$, with
\begin{align}
\mathcal{B}(B \to K^{*} Z') &= \tau_B\frac{|\vec{p}_{K^*}|}{4\pi}\bigg\{\Big|C_{bs}^V \tilde{V} - C_{bs}^T \tilde{T}_1\Big|^2 \nonumber \\
&\hspace{-4.5em} + \Big|C_{bs}^V \tilde{A}_1 - C_{bs}^T \tilde{T}_2\Big|^2 + 8\Big|C_{bs}^V \tilde{A}_{12} - C_{bs}^T \tilde{T}_{23}\Big|^2\bigg\}\,,
\label{eq:BtoKsZp}
\end{align}
where we define
\begin{align}
&\tilde{V} = \frac{|\vec{p}_{K^*}|}{m_B + m_{K^*}}V \,, &&\tilde{T}_{1} = 2|\vec{p}_{K^*}|T_1 \,,  \nonumber \\
&\tilde{A}_{1} = \frac{m_B + m_{K^*}}{2m_B} A_1 \,, &&\tilde{T}_2 = \frac{m_B^2 - m_{K^*}^2}{m_B}T_2 \,,\nonumber \\
&\tilde{A}_{12} = \frac{m_{K^*}}{M_{Z'}}A_{12} \,, &&\tilde{T}_{23} = \frac{M_{Z'}m_{K^*}}{m_B + m_{K^*}}T_{23}\,.
\end{align}
Here $V$, $A_1$, $A_{12}$, $T_1$, $T_2$ and $T_{23}$ are vector, axial-vector and tensor $B \to K^*$ transition form factors, respectively, evaluated at $q^2 = M_{Z'}^2$. The first two terms and the last term in Eq.~\eqref{eq:BtoKsZp} contribute to transversely and longitudinally polarised $K^*$ in the final state, respectively. In this work, we use the BSZ~\cite{Bharucha:2015bzk} parametrisation fit results of Ref.~\cite{Gubernari:2023puw} for the form factors in Eqs.~\eqref{eq:BtoKZp} and~\eqref{eq:BtoKsZp}.

\begin{figure}[t!]
  \centering
  \includegraphics[width=0.33\columnwidth]{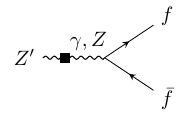}
  \includegraphics[width=0.32\columnwidth]{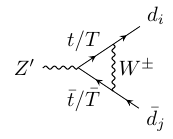}
  \includegraphics[width=0.3\columnwidth]{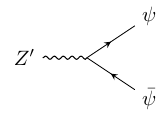}
  \caption{(Left and centre) One-loop induced decays of the light $Z'$ in the minimal  model. (Right) Tree-level decay of $Z'$ to a vector-like dark fermion $\psi$, considered in Sec.~\ref{sec:minimal_DM}.}
  \label{fig:Zp_decays}
\end{figure}

In order to contribute to the observed excess in $B^+\to K^+E_{\text{miss}}$, the $Z'$ must decay outside the Belle~II detector or to invisible final states. This occurs with the probability
\begin{align}
\label{eq:Pinv}
P_{\text{inv}} = P_{\text{out}} + (1 - P_{\text{out}})\mathcal{B}(Z'\to \text{inv.})\,,
\end{align}
where $P_{\text{out}} = \exp(-L/L_{Z'})$ and $L_{Z'} = \beta\gamma\tau_{Z'} = \beta\gamma/\Gamma_{Z'}$. Here, $\Gamma_{Z'}$ is the total decay width of $Z'$, $\mathcal{B}(Z'\to \text{inv.})$ is the invisible $Z'$ branching fraction, $L \approx 5~\text{m}$ is the distance from the interaction point to the edge of the Belle~II detector and $\beta\gamma$ is the boost of $Z'$ in the lab frame. The boost can be estimated to lie in the range $\beta\gamma \in [0.6,1.6]$ for $M_{Z'} = 2.1$~GeV and the incoming $e^+e^-$ energies at SuperKEKB~\cite{Ohnishi:2013fma}.

The $Z'$ decay width and invisible branching fraction are determined in the minimal model as follows. The total width of $Z'$ is $\Gamma_{Z'} =  3\Gamma_{\nu\bar{\nu}} + \Gamma_{e^+e^-} + \Gamma_{\mu^+\mu^-} + \Gamma_{\text{hadr.}}$, where
the $Z'$ decays to SM fermions via the kinetic and mass mixing, as shown in Fig.~\ref{fig:Zp_decays} (left), have the decay rate
\begin{align}
\label{eq:Zptoff}
\Gamma_{f_i \bar{f}_j} = \frac{M_{Z'}}{12\pi} \big[(v_{Z'}^f)^2 + (a_{Z'}^f)^2\big]_{ij} \,,
\end{align}
where $v_{Z'}^f$ and $a_{Z'}^f$ are given in Eq.~\eqref{eq:Zp_vector_axialvector}, and we neglect final-state masses. For $M_{Z'}=2.1$\,GeV the approximation is valid for $f=\nu,e,\mu$. For $M_{Z'}$ sufficiently above the regime of non-perturbative QCD and away from known hadronic thresholds, quark-hadron duality can be used to estimate the inclusive hadronic decay rate as
\begin{align}
\label{eq:Gamma_hadr}
\Gamma_{\text{hadr.}} = N_c \bigg(1 + \frac{\Delta_V}{\pi}\bigg)\bigg[\Theta_{u u}\Gamma_{u\bar u} + \hspace{-0.5em}\sum_{d_i,d_j = d,s}\hspace{-0.5em} \Theta_{d_i d_j}\Gamma_{d_i \bar d_j}\bigg] \,,
\end{align}
where $N_c = 3$ and $\Delta_V = \alpha_s + 0.522 \alpha^2_s - 1.04 \alpha_s^3  - 3.45 \alpha_s^4$ accounts for perturbative QCD corrections~\cite{Chetyrkin:1996ela,Baikov:2012er,Freitas:2014hra}, with $\alpha_s(M_{Z'}) = 0.297$ and $n_f = 3$ quark flavours. Down-type quark contributions to the inclusive hadronic decay rate are also generated by the vector coupling $C_{ij}^V$ and dipole-like coupling $C_{ij}^T$, as shown in Fig.~\ref{fig:Zp_decays} (centre). The former can be included by replacing $[v_{Z'}^d]_{ij} \to [v_{Z'}^d]_{ij} + C_{ij}^V/2$ and $[a_{Z'}^d]_{ij} \to [a_{Z'}^d]_{ij} + C_{ij}^V/2$ in Eq.~\eqref{eq:Zptoff}, while the latter contribution is proportional to the light quark masses and is therefore neglected. We include the kinematical factors $\Theta$ in Eq.~\eqref{eq:Gamma_hadr}, where $\Theta_{ij}^2 = 1 - (m_{P_i} + m_{P_j})^2/M_{Z'}^2$, to approximate the suppression of the decay rate due to the lightest hadronic thresholds ($Z'\to \pi\pi, \pi K, K K$), where $m_{P_i} = m_{\pi}$ is taken for $i = u,d$ and $m_{P_i} = m_{K}$ for $i = s$. We note that the applicability of this approach for $M_{Z'} = 2.1$~GeV is marginal but nevertheless sufficient for our purposes, as explained in more detail later in this section and in Sec.~\ref{sec:minimal_DM}.\footnote{A data driven approach using $e^+e^- \to \text{hadrons}$ and hadronic $\tau$ decay data would be necessary for $M_{Z'}\lesssim 2$~GeV to account for the effects of vector and axial-vector meson resonances ($\rho$, $\omega$, $\phi$, and $f_1$), as performed in Refs.~\cite{Ilten:2018crw,Bauer:2018onh,Baruch:2022esd,Foguel:2022ppx}.}

\begin{figure}[t!]
  \flushright
  \includegraphics[width=\columnwidth]{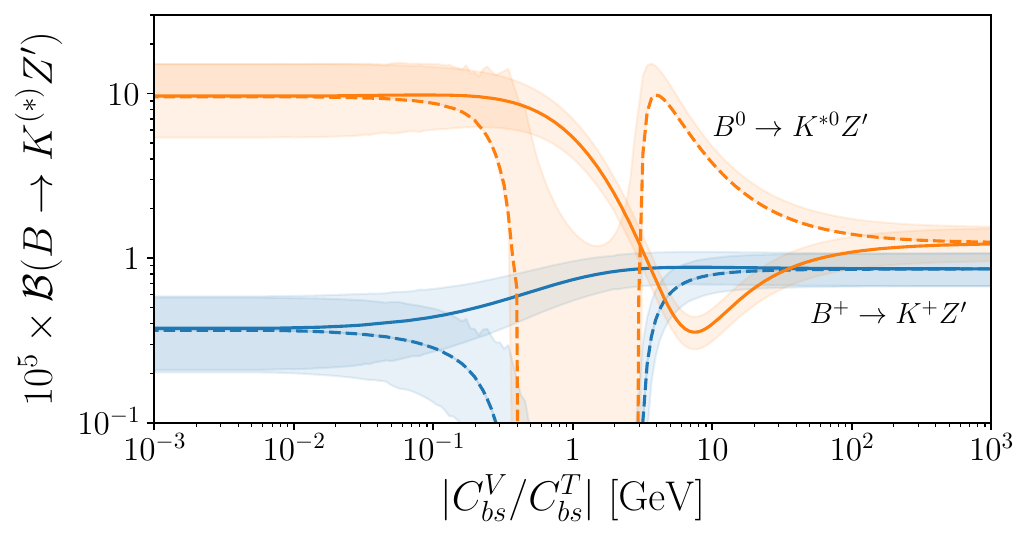}
  \caption{Favoured branching fractions for $B^+\to K^+Z'$ and $B^0\to K^{*0}Z'$ from the Belle~II~\cite{Belle-II:2023esi} and BaBar~\cite{BaBar:2013npw} data, as a function of the ratio $|C_{bs}^{V}/C_{bs}^{T}|$ for $M_{Z'} = 2.1$~GeV. Solid (dashed) lines show positive (negative) values of $C_{bs}^{V}/C_{bs}^{T}$.}
  \label{fig:fit_results}
\end{figure}

Only $Z'\to\nu\bar{\nu}$ can contribute to invisible $Z'$ decays in the minimal model, assuming that the other decay modes ($Z' \to e^+e^-/\mu^+\mu^-/\text{hadr.}$) are visible at Belle~II (in the central drift chamber, electromagnetic calorimeter or $K_L$–muon detector), yielding the invisible branching fraction $\mathcal{B}(Z'\to\text{inv.}) = 3\Gamma_{\nu\bar{\nu}}/\Gamma_{Z'}$.  The contribution of $Z'$ to the $B^+\to K^+ E_{\text{miss}}$ rate is then
\begin{align}
\label{eq:BtoKEmiss_tot}
\mathcal{B}(B^+ \to K^+ E_{\text{miss}}) &= \int dq^2\frac{d\mathcal{B}(B^+ \to K^+ \nu\bar{\nu})|_{\text{SM}}}{dq^2} \nonumber \\
&\hspace{1.3em} + \mathcal{B}(B^+ \to K^+ Z')P_{\text{inv}} \,,
\end{align}
where the factorisation of the second term holds in the narrow width approximation of $Z'$. The narrow width of $Z'$ also makes it safe to neglect the interference between the SM and $B^+ \to K^+(Z'\to) \nu\bar{\nu}$.

In Ref.~\cite{Belle-II:2023esi}, the Belle~II inclusive tagging analysis (ITA) search for $B^+ 
\to K^+ \nu\bar{\nu}$ observed an excess of events at the reconstructed momentum transfer $q_{\text{rec}}^2 \sim 4~\text{GeV}^2$. If the missing energy excess is interpreted as neutrinos (SM) plus the contribution of an undetected $Z'$ in the final-state, $B^+ \to K^+ Z'$, the ITA $q_{\text{rec}}^2$ spectrum data provides the best-fit mass $M_{Z'} = (2.1 \pm 0.1)$~GeV~\cite{Bolton:2024egx,Bolton:2025fsq}. The excess can then be interpreted as the presence of a vector and/or dipole-like coupling of the $Z'$ to the $b\to s$ transition. In Refs.~\cite{Bolton:2024egx,Bolton:2025fsq}, we considered the presence of one of these couplings at a time, performing a fit to the Belle~II data to find the best-fit coupling and therefore the implied size of $\mathcal{B}(B^+\to K^+ Z')$ in each case. As both couplings also contribute to the process $B\to K^* Z'$, the fit in Refs.~\cite{Bolton:2024egx,Bolton:2025fsq} also included the BaBar $q^2$ data used to derive an upper bound on $B^{0}\to K^{*0}\nu\bar{\nu}$~\cite{BaBar:2013npw}.\footnote{Very recently, Belle II published the results of their first search for inclusive $B \to X_s \nu \bar \nu$ decays~\cite{Belle-II:2025bho} using a sum over exclusive states. However their derived upper bound on $\mathcal B(B \to X_s \nu \bar \nu) < 3.2 \times 10^{-4}$ is not competitive with existing constraints targeting individual exclusive decay modes.}

Here, assuming that both the vector and dipole-like couplings are non-zero, we perform a fit to the Belle~II and BaBar $q^2$ data for $M_{Z'} = 2.1$~GeV, but now vary the ratio $r_c \equiv C_{bs}^{V}/C_{bs}^T$. From the fit, we obtain the expected branching fractions for $B^+\to K^+Z'$ and $B^0\to K^{*0}Z'$ shown in Fig.~\ref{fig:fit_results}. Firstly, when the vector or dipole-like coupling dominates, which occurs when $r_c$ is much above or below $\tilde{f}_T/f_+ \approx 1.5$~GeV, respectively, we recover the results of Ref.~\cite{Bolton:2024egx,Bolton:2025fsq}, i.e.
\begin{align}
\label{eq:fit}
\frac{\mathcal{B}(B^+\to K^+ Z')|_{\text{exp}}}{10^{-6}} = 
\begin{cases}
8.6_{-1.9}^{+2.1} & |r_c|\gg \tilde{f}_T/f_+\\
3.7_{-1.6}^{+2.1} & |r_c|\ll \tilde{f}_T/f_+
\end{cases}\,,
\end{align}
which can be seen to the far right and left of Fig.~\ref{fig:fit_results}. For a dominant dipole-like coupling, a smaller $\mathcal{B}(B^+\to K^+ Z')$ provides a better fit to the BaBar $q^2$ data, because $C_{bs}^T$ contributes more to $B^0\to K^{*0}Z'$ than $C_{bs}^V$. Explicitly,
\begin{align}
\label{eq:BtoK_vs_BtoKs}
\frac{\mathcal{B}(B^0\to K^{*0}Z')}{\mathcal{B}(B^+\to K^+ Z')} = 
\begin{cases}
1.43 & |r_c|\gg \tilde{f}_T/f_+\\
26.0 & |r_c|\ll \tilde{f}_T/f_+
\end{cases}\,.
\end{align}
For intermediate values of $r_c$, interference effects become important and depend on the sign of $r_c$. In Fig.~\ref{fig:fit_results}, the solid (dashed) lines illustrate the fit results for positive (negative) values of $r_c$. For $r_c < 0$, the terms in Eq.~\eqref{eq:BtoKZp} contributing to $B^+\to K^+Z'$ can interfere destructively, and for $-3\lesssim r_c/\text{GeV} \lesssim -0.04$, it is no longer possible to address the Belle~II excess and be compatible with the BaBar $q^2$ data. For $r_c > 0$, destructive interference may instead occur in $\mathcal{B}(B^0\to K^{*0}Z')$, which can be seen for $4\lesssim r_c/\text{GeV} \lesssim 20$ in Fig.~\ref{fig:fit_results}.

For the minimal aligned $U(1)'$ model to accommodate the Belle~II excess, it must satisfy 
\begin{align}
\label{eq:BelleII_constraint}
\mathcal{B}(B^+ \to K^+ Z')P_{\text{inv}} = \mathcal{B}(B^+\to K^+ Z')|_{\text{exp}}\,,
\end{align}
where the left-hand side is the model prediction and the right-hand side is the best-fit branching fraction from the Belle~II and BaBar $q^2$ data. However, we find that when the combination $|C_{bs}^{V} f_+ + C_{bs}^T \tilde{f}_T |$ in Eq.~\eqref{eq:BtoKZp} is large enough to saturate Eq.~\eqref{eq:BelleII_constraint}, the vector coupling always dominates over the dipole-like coupling. Specifically, $C_{bs}^V$ is determined at the scale $\mu = M_W$ predominantly by the direct coupling of $Z'$ to $t/T$ and $C_{bs}^T$ by the $\gamma-Z'$ mixing. The ratio of couplings is then given at $\mu = M_W$ by
\begin{align}
\label{eq:CV_versus_CT}
\hspace{-0.5em}\,\frac{|r_c|}{\text{GeV}} \approx 2\times 10^4~\bigg[\frac{y_T^t}{0.1}\bigg]^2\bigg[\frac{\text{TeV}}{m_T}\bigg]^2\bigg[\frac{10^{-3}}{\tilde{g}q'}\bigg]^2\bigg[\frac{X_t/L_T}{0.07}\bigg]\,.
\end{align}
Running under QCD to $\mu \sim m_b$ is expected to modify the numerical size of the dipole-like coupling at most by a factor of a few, and thus the large hierarchy between the couplings as expressed in Eq.~\eqref{eq:CV_versus_CT} remains valid at $\mu \sim m_b$. In Eq.~\eqref{eq:BelleII_constraint}, we therefore neglect the dipole-like coupling and use the best-fit branching fraction in the limit $|r_c|\gg \tilde{f}_T/f_+$, which is equivalent to requiring $|C_{bs}^V|^2 P_{\text{inv}} = |C_{bs}^V|^2|_{\text{exp}}$ with the best-fit vector coupling $ C_{bs}^V|_{\text{exp}}= (1.4 \pm 0.5)\times 10^{-8}$~\cite{Bolton:2024egx}.

\begin{figure}[t!]
  \flushright
  \includegraphics[trim={0 1.2cm 0 0},clip,width=0.997\columnwidth]{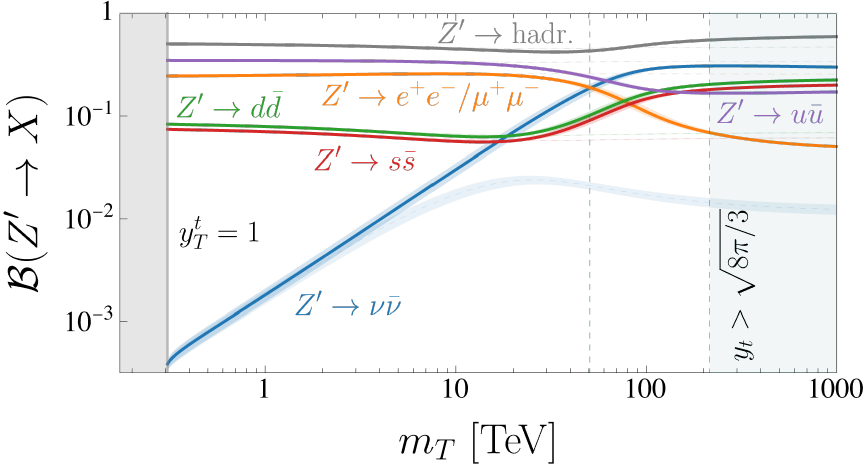}
  \includegraphics[width=\columnwidth]{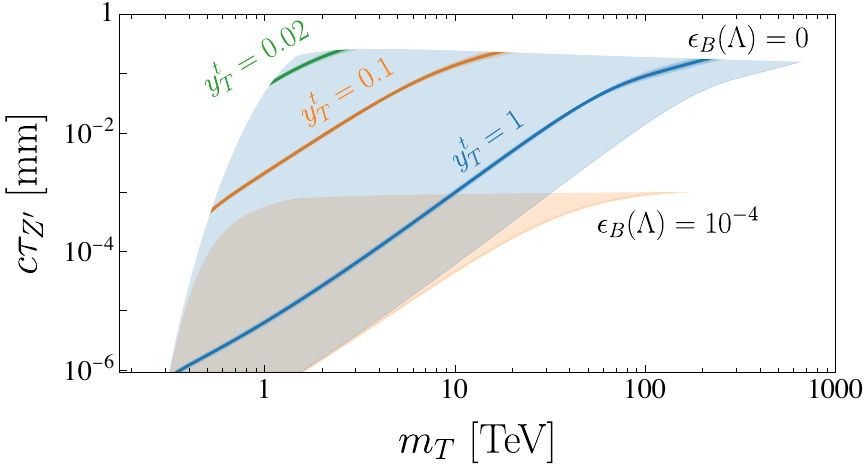}
  \caption{(Above) Branching fractions of $Z'$ in the minimal model accommodating the Belle II measurement in Eq.~\eqref{eq:BelleII_constraint} as a function of $m_T$, for $y_T^t = 1$. Solid (faint dashed) lines are shown for $\epsilon_{B}(\Lambda) = 0$ ($10^{-4}$). Also shown are $y_t$ perturbativity limits (vertical dashed lines) for both cases. (Below) Proper lifetime of $Z'$ as a function of $m_T$ for $\epsilon_{B}(\Lambda) = 0$ and different values of $y_T^t$. Envelopes of possible $c\tau_{Z'}$ values are shown for $\epsilon_{B}(\Lambda) = 0$ and $10^{-4}$.}
  \label{fig:BR_plot_Zprime}
\end{figure}

In Fig.~\ref{fig:BR_plot_Zprime}, we investigate the properties of $Z'$ if Eq.~\eqref{eq:BelleII_constraint} is satisfied and $B^+ \to K^+(Z'\to) \nu\bar{\nu}$ therefore provides the excess in missing energy at Belle~II. Above, we show the branching fractions of $Z'$ as a function of $m_T$ for $y_T^t = 1$. The solid (dashed) lines show the choice of the kinetic mixing at the high scale $\epsilon_B(\Lambda) = 0$ ($10^{-4}$), while the bands indicate the $\pm 1\sigma$ uncertainty from the best-fit branching fraction in Eq.~\eqref{eq:BelleII_constraint}. We show the inclusive hadronic branching fraction $Z' \to \text{hadr.}$ and its contributions from different light quark flavours ($ \Gamma_{u\bar{u}/d\bar{d}/s\bar{s}}$ induced by $\epsilon_{A,Z}$ and $\delta_Z$ dominate, while $\Gamma_{ d\bar{s}/s\bar{d}}$ from $C_{ij}^V$ are negligible) and the leptonic branching fractions $Z'\to\nu\bar{\nu}/e^+e^-/\mu^+\mu^-$. We note that for $y_T^t = 1$, it is not possible to satisfy Eq.~\eqref{eq:BelleII_constraint} for $m_T \lesssim 300~\text{GeV}$, due to the suppression of $\mathcal{B}(Z' \to \nu\bar{\nu})$, which we indicate as a dark gray region.  The observed branching fractions follow purely from the behaviour of the kinetic and mass mixing. For $m_T \lesssim 10~\text{TeV}$, $\epsilon_A\gg\Delta_{Z'}$ and $Z'$ behaves like a dark photon with severely suppressed couplings to SM neutrinos in both $\epsilon_B(\Lambda) = 0$ ($10^{-4}$) scenarios. 
For $\epsilon_B(\Lambda) = 0$ and $m_T \gtrsim 10~\text{TeV}$, $\epsilon_Z\ll\tilde\delta_A$ and $Z'$ has SM $Z$-like branching fractions. In this limit, $\mathcal{B}(Z' \to\nu\bar{\nu}) = 0.29$. 
On the other hand, in the $\epsilon_B(\Lambda) = 10^{-4}$ scenario, the additional contribution to $\epsilon_A$ suppresses the branching fractions to neutrinos in the whole considered $m_T$ range. When $m_T \gtrsim 200$~TeV ($50$~TeV), the Yukawa coupling $y_t$ violates perturbative unitarity for $\epsilon_B(\Lambda) = 0$ ($10^{-4}$), indicated as a vertical green dashed line and shaded (unshaded) region. Direct experimental probes of $y_t$ discussed in Sec.~\ref{subsubsec:collider} represent even stronger constraints. 
Finally note that, for any given value of $m_T$ there is a (fine-tuned) value of $\epsilon_B(\Lambda)$ that exactly cancels the value of $\epsilon_A$ induced by the running, leading to the $Z'$ branching fractions that are SM $Z$-like. In the following we do not consider such fine-tuned solutions.

In Fig.~\ref{fig:BR_plot_Zprime} (below), we instead show the expected proper lifetime of $Z'$ ($c\tau_{Z'}$) as a function of $m_T$, for $\epsilon_B(\Lambda) = 0$ and three $y_T^t$ values. As a blue (orange) envelope we also illustrate the total range of possible $c\tau_{Z'}$ values for $\epsilon_B(\Lambda) = 0$ ($10^{-4}$); the upper and lower edges correspond to the maximum values of $y_t$ and $y_T^t$ compatible with perturbative unitarity, respectively. On the left edges of the envelopes, the Belle~II data cannot be satisfied due to the suppression of $\mathcal{B}(Z'\to \nu\bar{\nu})$. In general, it can be seen that the lifetime of $Z'$ cannot exceed $c\tau_{Z'}\sim 0.3~\text{mm}$, because the lifetime is maximised when the irreducible mass mixing dominates the interactions of $Z'$. For larger $\epsilon_{B}(\Lambda)$, in the absence of fine-tuning, $Z'$ is more dark photon-like and decays faster. For the expected boosts at Belle~II, we conclude that the probability of $Z'$ decaying outside the detector is negligible, so $P_{\text{inv}} \approx \mathcal{B}(Z'\to \nu\bar{\nu})$. Thus, the effective size of the Belle~II detector, $L\approx 5$~m, has no impact on the results.

Finally, we note that the consistency condition in Eq.~\eqref{eq:consistency} implies that there is a minimum value of $y_T^t$ which is compatible with the Belle~II data. If Eq.~\eqref{eq:consistency} is saturated, the vector coupling $C_{bs}^V$ is given by Eq.~\eqref{eq:CV_limit}. In this limit $|C_{bs}^V|^2 P_{\text{inv}} = |C_{bs}^V|^2|_{\text{exp}}$ can be rearranged for $y_T^t$, giving
\begin{align}
\label{eq:yTt_lower}
y_T^t \gtrsim \frac{\sqrt{2}C_{bs}^V|_{\text{exp}}}{m_t M_{Z'} \mathcal{C}_{bs}\sqrt{P_{\text{inv}}}}\approx 0.012\bigg[\frac{0.29}{P_{\text{inv}}}\bigg]^{1/2}\,,
\end{align}
for $m_t \ll m_T$, $\epsilon_B(\Lambda) = 0$, and $P_{\text{inv}}$ evaluated at the consistency limit of Eq.~\eqref{eq:consistency}. In the minimal model, the $Z'$ interactions are SM $Z$-like in this regime, and thus $P_{\text{inv}} = 0.29$. The lower bound on $y_T^t$ can be seen visually in Fig.~\ref{fig:BR_plot_Zprime} in the top left corner of the $c\tau_{Z'}$ envelope for $\epsilon_B(\Lambda) = 0$. Using Eq.~\eqref{eq:consistency}, it is also possible to derive a lower bound on the gauge coupling, $\tilde{g}q' \gtrsim C_{bs}^V|_{\text{exp}}/\mathcal{C}_{bs}m_t m_t\sqrt{P_{\text{inv}}}$.

In Fig.~\ref{fig:BR_plot_Zprime} (above), it can be seen that the $Z'$ also has sizeable branching fractions to $e^+e^-/\mu^+\mu^-$ in both the dark photon and SM $Z$-like limits. Stringent constraints therefore apply from measurements of $B\to K^{(*)}\ell^+\ell^-$, with $\ell = e,\mu$ being either prompt or displaced depending on the value of $c\tau_{Z'}$. For prompt $Z' \to e^+ e^-/\mu^+\mu^-$ decays, the $Z'$ will appear as a peak of events in the $[q_i^2,q_j^2]$ bin around $q^2 = M_{Z'}^2$~\cite{LHCb:2014cxe,BELLE:2019xld}. 
On the other hand, dedicated LHCb searches for $B$ decays to resonant displaced dimuons provide even stronger bounds on the process $B\to K^{(*)}(Z'\to)\mu^+\mu^-$~\cite{LHCb:2015nkv,LHCb:2016awg}. For $Z'$ lifetimes in the range $c\tau_{Z'} \in[0.03,300]$~mm, the searches place an upper limit on $\mathcal{B}(B \to K^{(*)} Z')\mathcal{B}_{\mu^+\mu^-}$, with $\mathcal{B}_{\mu^+\mu^-} = \mathcal{B}(Z'\to\mu^+\mu^-)$. However, even for $c\tau_{Z'} < 0.03$~mm, when the $Z'$ decays within the decay time resolution of the experiment, the limits for $c\tau_{Z'} = 0.03$~mm remain applicable.

\begin{figure}[t!]
  \flushright
  \includegraphics[width=\columnwidth]{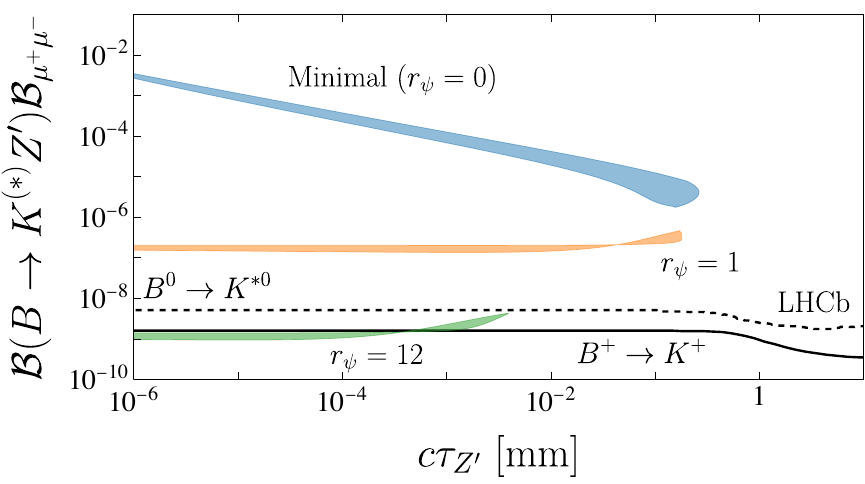}
  \caption{The branching fractions of $B\to K^{(*)} (Z'\to)\mu^+\mu^-$ in the model with a vector-like dark fermion $\psi$ (for two values of the $U(1)'$ charge ratio $r_\psi$) which accommodate the Belle~II data, compared to the LHCb upper limits (black, solid for $B\to K$ and dashed for $B\to K^*$). See text for details.}
  \label{fig:LHCb}
\end{figure}

In Fig.~\ref{fig:LHCb}, we demonstrate how the LHCb constraints exclude the minimal $U(1)'$ model as an explanation of the Belle~II excess. The black solid (dashed) line shows the 95\%~CL upper bound on the $B\to K^{(*)}(Z'\to)\mu^+\mu^-$ branching fraction from LHCb. The blue region shows the expected range of $\mathcal{B}(B\to K^{(*)}Z')\mathcal{B}_{\mu^+\mu^-}$ values if the Belle~II data is accommodated, for $\epsilon_B(\Lambda) = 0$. The observed upper limit and expected branching fraction of the $B^0\to K^{*0}$ mode are rescaled by the ratio in Eq.~\eqref{eq:BtoK_vs_BtoKs} in the limit $|r_c|\gg \tilde{f}_T/f_+$, so that the expected branching fractions of the $B^+\to K^{+}$ and $B^0\to K^{*0}$ modes coincide in the figure. It is evident that the LHCb data exclude all of the viable parameter space of the minimal model. Furthermore, different values of $\epsilon_B(\Lambda)$ do not change this conclusion; even if fine-tuning suppresses $\epsilon_A$, the $Z'$ still decays with a large branching fraction to $Z'\to\mu^+\mu^-$ via the mass mixing.\footnote{This bound could thus also be important for the gauged $U(1)_{B_3-L_3}$ model proposed in Ref.~\cite{Altmannshofer:2023hkn}, even though the $Z'$ width is dominated by the tree-level mediated $Z' \to \nu_\tau \bar \nu_\tau$.}

We briefly return to the method used to estimate the inclusive hadronic decay rate in Eq.~\eqref{eq:Gamma_hadr}. Uncertainties from the use of perturbative QCD corrections instead of the data driven approach enter the estimate of the total $Z'$ width and therefore all $Z'$ branching fractions. While these uncertainties may result in small changes to the results presented in Figs.~\ref{fig:BR_plot_Zprime} and~\ref{fig:LHCb}, the decisive exclusion of the model by LHCb makes the study of such uncertainties unnecessary.

To finish this section, we conclude that in order for the $U(1)'$ gauge boson $Z'$, coupled to the top partner $T'$, to provide the missing energy signature and evade the LHCb bounds, it must decay invisibly to additional light, non-SM states. It is plausible that there are SM-singlet fields charged under the $U(1)'$. Such feebly-interacting states would naturally evade detection and could make up an extended dark sector. In Sec.~\ref{sec:minimal_DM}, we consider the scenario of a single vector-like dark fermion $\psi$ with a $U(1)'$ charge. First, we review how this addition changes the phenomenology at Belle~II, and explore how other experimental probes (in addition to LHCb) constrain the parameter space. In this set up, we then explore the interesting possibility of $\psi$ being some or all of the DM.

%
\section{Model with Dark Fermion}
\label{sec:minimal_DM}
%

Here we extend the minimal model of Sec.~\ref{sec:minimal}, by introducing a SM-singlet vector-like dark fermion $\psi({\bf 1},{\bf 1},0,q_\psi')$ which we allow to have a different $U(1)'$ charge ($q_\psi'$)  compared to $T'$ and $\Phi$. The Lagrangian now includes 
\begin{align}
\label{eq:model_DM}
\mathcal{L} &\supset \bar{\psi}(i\slashed{D} - m_\psi)\psi \,.
\end{align}
The dark fermion $\psi$ interacts with $Z'$ and $Z$ according to Eqs.~\eqref{eq:Zp_vector_axialvector}~and~\eqref{eq:Z_vector_axialvector}, respectively, with $\mathfrak{v}^\psi = q_\psi'$, $\mathfrak{a}^\psi = 0$. There are two additional free parameters with respect to the minimal $U(1)'$ model: the vector-like mass $m_\psi$ and the $U(1)'$ charge ratio $r_\psi = q_\psi'/q'$.

\subsection{Phenomenology in $B$ Decays}
\label{subsec:belleii_DM}

At Belle~II, the production of $Z'$ with $M_{Z'} = 2.1$~GeV is the same as described in Sec.~\ref{sec:minimal}, i.e. $B^+\to K^+Z'$ in Fig.~\ref{fig:BtoKZprime}. For $2m_\psi > M_{Z'}$, the $Z'$ cannot decay at tree-level via $Z'\to \psi\bar{\psi}$ (shown in Fig.~\ref{fig:Zp_decays}, right) and the discussion is identical to Sec.~\ref{sec:minimal}. For $2m_\psi < M_{Z'}$ on the other hand, $Z'\to \psi\bar{\psi}$ becomes kinematically allowed and the total decay width is given by $\Gamma_{Z'} =  3\Gamma_{\nu\bar{\nu}} + \Gamma_{e^+e^-} + \Gamma_{\mu^+\mu^-} + \Gamma_{\text{hadr.}} + \Gamma_{\psi\bar{\psi}}$, with the decay rate for $Z'\to \psi\bar{\psi}$,
\begin{align}
\label{eq:ZptoXX}
\Gamma_{\psi\bar{\psi}} = \frac{M_{Z'}}{12\pi} f_\psi (\tilde{g} q_\psi')^2 \,.
\end{align}
Here, $f_\psi = f(m_\psi^2/M_{Z'}^2)$ with $f(x) = (1 + 2 x)\sqrt{1 - 4 x}$. In addition to the three neutrino final states, $Z' \to \psi\bar{\psi}$ also contributes to the total invisible branching fraction as $\mathcal{B}(Z'\to \text{inv.}) = (3\Gamma_{\nu\bar{\nu}}+\Gamma_{\psi\bar{\psi}})/\Gamma_{Z'}$.

\begin{figure}[t!]
  \centering
  \includegraphics[trim={0 1.2cm 0 0},clip,width=\columnwidth]{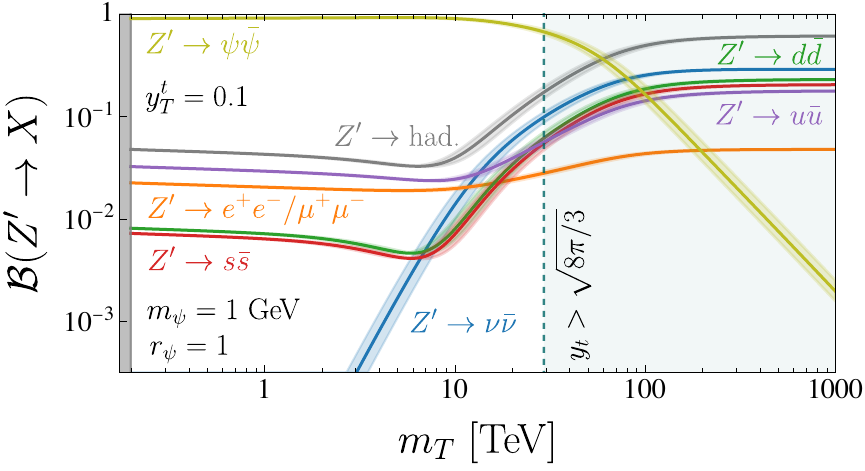}
  \includegraphics[width=\columnwidth]{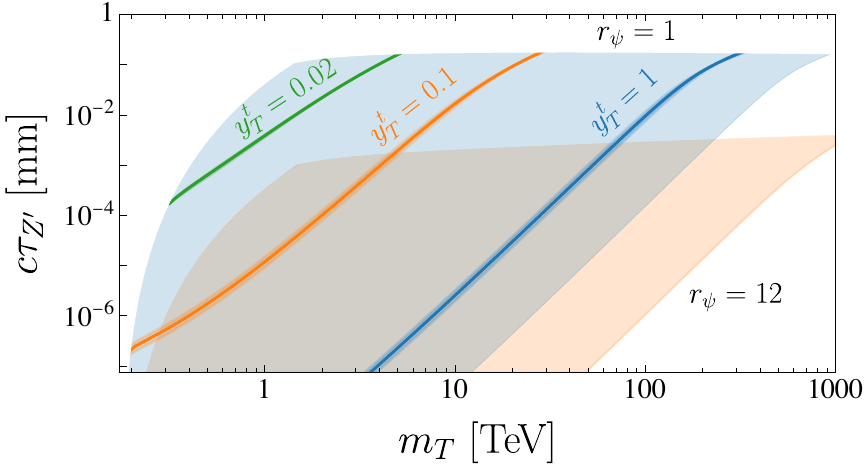}
  \caption{As in Fig.~\ref{fig:BR_plot_Zprime}, but with branching fractions (above) and the proper lifetime (below) of $Z'$ in the model with a dark fermion $\psi$, for $y_T^t = 0.1$, $m_\psi = 1~\text{GeV}$ and $r_\psi = 1$. The envelope of possible $c\tau_{Z'}$ values is also shown for $r_\psi = 12$.}
  \label{fig:BR_plot_Zprime-2}
\end{figure}

We now repeat the exercise of the previous section and determine where Eq.~\eqref{eq:BelleII_constraint} holds, and therefore the Belle~II excess is accommodated in the extended model. We furthermore examine the properties of $Z'$, $T$ and $\psi$ in the viable parameter space. For simplicity in the following we restrict our discussion to the limit $\epsilon_B(\Lambda)=0$. Firstly, in Fig.~\ref{fig:BR_plot_Zprime-2} (above), we show the $Z'$ branching fractions as a function of $m_T$ for $y_T^t = 0.1$, $m_\psi = 1$~GeV and $r_\psi = 1$. We first note the dominance of the decay mode $Z'\to \psi\bar{\psi}$ for $m_T\lesssim 7$~TeV. In this regime the $Z'$ is again dark photon-like, with $Z'\to \psi\bar{\psi}$ making up for the shortfall in $Z'\to \nu\bar{\nu}$ and the other decay modes being suppressed with respect to Fig.~\ref{fig:BR_plot_Zprime}. As $r_\psi$ is increased, the other decay modes are further suppressed relative to $Z'\to \psi\bar{\psi}$. For $m_T\gtrsim 7$~TeV, the branching fractions again tend towards the SM $Z$-like limit. Eventually however, for $m_T\gtrsim 30$~TeV, the Yukawa coupling $y_t$ violates perturbative unitarity. In Fig.~\ref{fig:BR_plot_Zprime-2} (below), we instead show $c\tau_{Z'}$ as a function of $m_T$ for $m_\psi = 1$~GeV, $r_\psi = 1$, and three values of $y_T^t$. The blue (orange) envelope shows the total range of possible $c\tau_{Z'}$ values for $r_\psi = 1$ ($12$). The presence of $Z'\to \psi\bar{\psi}$ extends the range of possible $c\tau_{Z'}$ values with respect to Fig.~\ref{fig:BR_plot_Zprime}. Larger values of $r_\psi$ increase the partial decay rate for $Z'\to \psi\bar{\psi}$ and therefore decrease the maximum $c\tau_{Z'}$. For $r_\psi\ll 1$, the $Z'$ branching fractions and lifetime tend towards those shown in Fig.~\ref{fig:BR_plot_Zprime}. Again, $c\tau_{Z'} \sim 0.3$~mm presents an absolute maximum possible lifetime of $Z'$. 

\begin{figure}[t!]
  \flushright
  \includegraphics[trim={0 1.2cm 0 0},clip,width=\columnwidth]{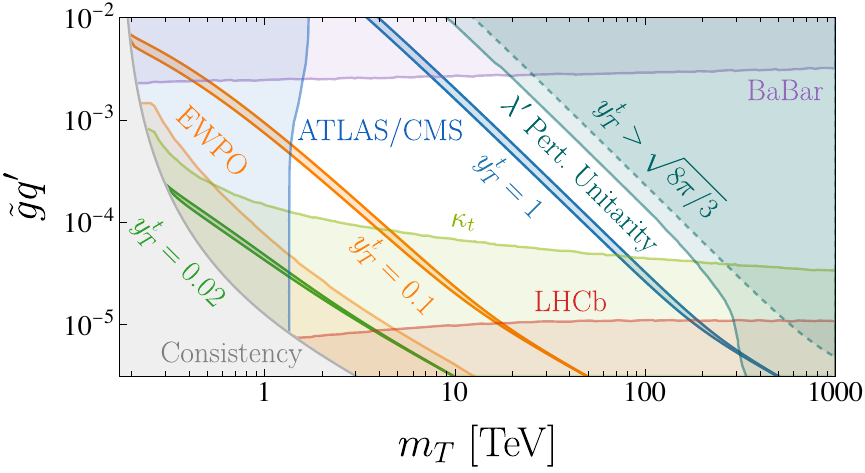}
  \includegraphics[width=\columnwidth]{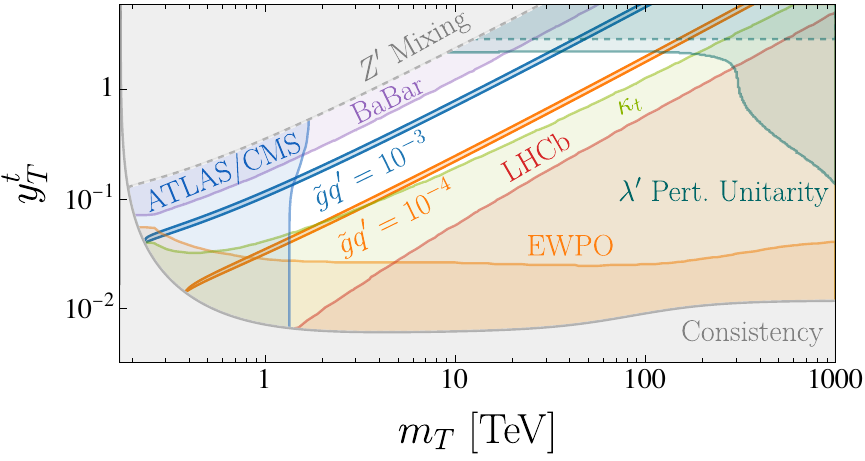}
  \caption{(Above) Allowed and excluded regions in the $(m_T,\tilde{g}q')$ plane for the $U(1)'$ charge ratio $r_\psi = 12$ and vector-like dark fermion mass $m_\psi = 1$~GeV. The Belle~II excess can be accommodated anywhere above the \textit{Consistency} region (in gray), with the blue, orange and green bands indicating the favoured regions for fixed $y_T^t$. Other constraints originate from dark photon searches at BaBar (purple), collider searches for top partners at ATLAS/CMS (blue), the measurements of the top Yukawa coupling ($\kappa_t$, light green), LHCb search for $B \to K^{(*)} (Z' \to \mu^+ \mu^-)$ (red), EWPO (orange), and perturbative unitarity (dark green). (Below) The same constraints, but in the $(m_T, y_T^t)$ plane. The Belle~II favoured regions are shown as blue and orange bands for two fixed values of $\tilde{g}q'$. See text for details.}
  \label{fig:gtqp_plot}
\end{figure}

With these results, we can return to whether the model is compatible with the displaced $B \to K^{(*)}(Z'\to)\mu^+\mu^-$ LHCb bounds. In Fig.~\ref{fig:BR_plot_Zprime-2} (above), it can be seen that the $Z'\to \mu^+\mu^-$ branching fraction remains appreciable for $r_\psi = 1$. Thus, the $U(1)'$ charge ratio $r_\psi$ must be large enough for $\mathcal{B}_{\mu^+\mu^-}$ to be sufficiently suppressed. In Fig.~\ref{fig:LHCb}, we show as orange and green regions the expected range of $\mathcal{B}(B\to K^{(*)}Z')\mathcal{B}_{\mu^+\mu^-}$ values for $r_\psi = 1$ and $12$, respectively. Comparing to the 95\%~CL upper limits, the $r_\psi = 1$ scenario is excluded, while $r_\psi = 12$ evades the $B\to K$ constraint for sufficiently small $c\tau_{Z'}$. An estimate of the required $r_\psi$ can be found using  $\mathcal{B}_{\mu^+\mu^-} \ll 1$ when $\mathcal{B}(Z'\to \psi\bar{\psi}) \approx 1$ and the other $Z'$ decay modes are dark photon-like but suppressed. This also applies to the inclusive hadronic decay modes, making the impact of uncertainties from the use of perturbative QCD estimates negligible. Then, Eqs.~\eqref{eq:BtoKZp}, \eqref{eq:Zptoff} and \eqref{eq:ZptoXX} can be combined to find that
\begin{align}
r_\psi \gtrsim 10.1 \bigg[\frac{0.44}{f_\psi}\bigg]^{1/2}\bigg[\frac{L_T}{55}\bigg]\,,
\end{align}
must be satisfied to evade the LHCb bounds. Note that the relatively large charge ratio required can also be accommodated by introducing several dark fermions with $\mathcal O(1)$ charge ratios. We briefly comment on this and other possible model extensions in Sec.~\ref{sec:conclusions}.

As the model is not ruled out by LHCb for $r_\psi \gtrsim 10$, we now examine other constraints on the model, which will be summarised in Sec.~\ref{subsec:pheno_DM}. In Fig.~\ref{fig:gtqp_plot}, we show the most stringent bounds in the $(m_T,\tilde{g}q')$ plane (above) and $(m_T,y_T^t)$ plane (below), for the benchmark parameters $r_\psi = 12$ and $m_\psi = 1$~GeV. The Belle~II excess can be accommodated anywhere above the gray region labelled \textit{Consistency}, which is inaccessible as it violates the consistency condition in Eq.~\eqref{eq:consistency}. On the other hand, the value of $y_T^t$ needed to explain the Belle~II data eventually violates perturbative unitarity for $y_T^t > \sqrt{8\pi/3}$, which we show as a dark green dashed line. We also show the regions of the parameter space which are favoured by the Belle~II data for specific values of $y_T^t$ (above) and $\tilde{g}q'$ (below), with the bands indicating the $\pm 1
\sigma$ uncertainty on $C_{bs}^V|_{\text{exp}}$. For $\tilde{g}q'$ well above the consistency bound in Eq.~\eqref{eq:consistency}, an estimate of the favoured region can be obtained by rearranging $|C_{bs}^V|^2 P_{\text{inv}} = |C_{bs}^V|^2|_{\text{exp}}$ using Eq.~\eqref{eq:CV} (neglecting the contribution from $Z-Z'$ mixing) and $P_{\text{inv}} \approx \mathcal{B}(Z'\to \psi\bar{\psi}) \approx 1$, to find
\begin{align}
\tilde{g}q' &\approx \frac{\mathcal{C}_{bs}}{C_{bs}^V|_{\text{exp}}}\bigg(\frac{m_t M_{Z'}y_T^t}{2m_T}\bigg)^2 X_t \nonumber \\
&\approx 1.2\times 10^{-3}\bigg[\frac{y_T^t}{0.5}\bigg]^2\bigg[\frac{10~\text{TeV}}{m_T}\bigg]^2\bigg[\frac{X_t}{8.6}\bigg] \,. \label{eq:BelleII_approx}
\end{align}
As $\tilde{g}q'$ approaches the consistency bound, the scaling changes to $\tilde{g}q'\approx M_{Z'}y_T^t/\sqrt{2}m_T$. On the other hand, this regime is excluded by the other existing experimental constraints. The region not yet excluded thus lies between $1.3\lesssim m_T/\text{TeV}\lesssim 170$ and $4\times 10^{-5}\lesssim \tilde{g}q'\lesssim 3\times 10^{-3}$. Converted to the vev of $\Phi$, we then find that the Belle~II excess can be viably explained for $0.8\lesssim \tilde{v}/\text{TeV}\lesssim 48$.

Before summarising the experimental bounds on the model in the next section, we describe the remaining theoretical bounds shown in Fig.~\ref{fig:gtqp_plot}. The parameter space where perturbative unitarity is violated by the quartic coupling $\lambda'$ in Eq.~\eqref{eq:lambda_prime} is shown as a dark green shaded region. 
This boundary of the allowed region in Fig.~\ref{fig:gtqp_plot} is given by the approximate condition
\begin{align}
\label{eq:pert_bound}
\tilde{g}q' \lesssim 8.6\times 10^{-3}\bigg[\frac{10~\text{TeV}}{m_T}\bigg]^2\bigg[\frac{55}{L_T}\bigg]\bigg[\frac{X_t}{8.6}\bigg]^2\,.
\end{align}

In the $(m_T, y_T^t)$ plane, we do not show the parameter space in the top left corner where $\tilde{g}q'> 10^{-2}$ is needed to accommodate the Belle~II data and thus the leading order expansion used for the $Z'$ mixing is no longer valid. In this regime, $Z-Z'$ mixing is also expected to dominate $C_{bs}^V$ in Eq.~\eqref{eq:CV} and therefore the $B^+\to K^+ Z'$ process. The scenario where $Z'$ is produced via the $Z-Z'$ mixing and decays dominantly via $Z'\to \psi\bar{\psi}$ was considered previously in Ref.~\cite{Calibbi:2025rpx}.\footnote{To evade the constraints on $\epsilon_A$ from dark photon searches at BaBar, discussed in the next section, the $Z-Z'$ mixing in this case must be induced principally by the mass mixing. The simplest model realisation is an $SU(2)_L$ doublet scalar $\Phi'$ charged under $U(1)'$, with $\mathcal{L}\supset |D_\mu \Phi'|^2$ generating tree-level mass mixing between $Z$ and $Z'$~\cite{Calibbi:2025rpx}.}

\subsection{Other Phenomenology}
\label{subsec:pheno_DM}

Here, we review the other constraints shown in Fig.~\ref{fig:gtqp_plot}. The strongest bounds come from direct ATLAS and CMS searches for top partners~\cite{ATLAS:2024gyc}, the CMS measurement of the top quark Yukawa coupling~\cite{CMS:2020mpn}, and $e^+e^-\to \gamma Z'$ followed by $Z'\to\text{inv.}$ at BaBar~\cite{BaBar:2017tiz}. We give a summary of these and other constraints in the next few subsections. 

\subsubsection{Dark Photon Searches}
\label{subsubsec:darkPhoton}

Direct searches for dark photons with $M_{Z'} = 2.1$~GeV have been performed at BaBar~\cite{BaBar:2014zli,BaBar:2017tiz}, BESIII~\cite{BESIII:2017fwv} and LHCb~\cite{LHCb:2017trq,LHCb:2019vmc}. The most stringent bounds on the kinetic mixing $\epsilon_A$ are derived from a search for $e^+e^- \to \gamma Z'$ followed by $Z'\to e^+e^-/\mu^+\mu^-$~\cite{BaBar:2014zli} or $Z'\to \text{inv.}$~\cite{BaBar:2017tiz} at BaBar. The former process constrains the minimal $U(1)'$ scenario in Sec.~\ref{sec:minimal} for large $\tilde g q' $. However, if $Z'\to \psi\bar{\psi}$ dominates the $Z'$ decay width to accommodate the Belle~II data, only the latter process is relevant in the parameter space of Fig.~\ref{fig:gtqp_plot}. We recast the bound as $\epsilon_A^2 \mathcal{B}(Z'\to \text{inv.}) < 7.8 \times 10^{-4}$ (90\%~CL), with the exclusion shown as a purple shaded region in Fig.~\ref{fig:gtqp_plot}. Taking $\epsilon_A = -\epsilon_Z/t_w = e\tilde{g}q'L_T/6\pi^2$ and $\mathcal{B}(Z'\to \text{inv.}) \approx 1$, we find that
\begin{align}
\label{eq:BaBar_bound}
\tilde{g}q' \gtrsim 2.7\times 10^{-3} \bigg[\frac{55}{L_T}\bigg]\,,
\end{align}
is approximately excluded by the BaBar data.

\subsubsection{Collider Searches}
\label{subsubsec:collider}

If light enough, the top partner $T$ can be pair-produced at the LHC, with the production cross section dominated by the QCD-induced $gg \to T\bar{T}$ partonic process. The final-state, and therefore the signature probed by ATLAS or CMS experiments, depends on how $T$ decays. 

\begin{figure}[t!]
  \flushright
  \includegraphics[width=\columnwidth]{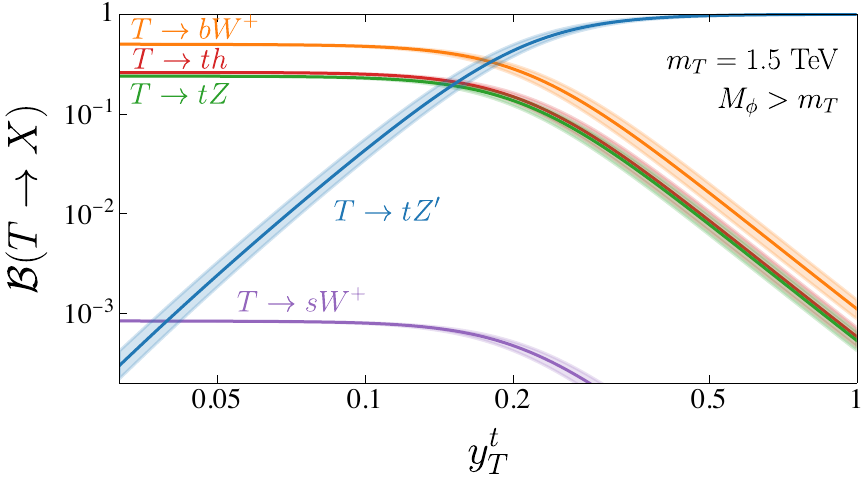}
  \caption{Branching fractions of $T$ satisfying the Belle~II data in the minimal aliged $U(1)'$ plus $\psi$ scenario, as a function of $y_T^t$ for $m_T  = 1.5~\text{TeV}$ and assuming $M_\phi > m_T$.}
  \label{fig:BR_plot_top}
\end{figure}

In Fig.~\ref{fig:BR_plot_top}, we illustrate the branching fractions of $T$ as a function of $y^t_T$, for $m_T = 1.5$~TeV and $M_{Z'} = 2.1$~GeV. For $y_T^t\lesssim 0.2$, the branching fractions tend towards those of a typical $SU(2)_L$ singlet top partner. For $m_t \ll m_T$, the decay rates are $\Gamma(T\to \text{SM}) \approx m_T^2s_L^2 C/32\pi v^2$ where $C \in \{2|V_{tb}|^2,1,1\}$ for $\text{SM} \in \{bW^+, t Z, t h\}$, predicting $\mathcal{B}(T\to \text{SM}) \approx \{1/2,1/4,1/4\}$. Instead, for $y_T^t\gtrsim 0.2$, the $Z'$ coupling dominates, with $\Gamma(T\to tZ') \approx m_T^2s_R^2/32\pi\tilde{v}^2$ and $\mathcal{B}(T\to t Z') \approx 1$. For simplicity, we assume that the scalar $\phi$ is heavier than $T$, forbidding $T\to t\phi$ decays. For $M_\phi \ll m_T$, $T\to tZ'/t\phi$ would have equal decay rates.

In the $SU(2)_L$ singlet-like limit, currently the most sensitive signature at the LHC is $pp \to T\bar{T}\to b \bar{b} W_l^\pm W_h^\mp$. Here, one final-state $W^\pm$ is required to decay leptonically, $W_l^\pm\to \ell^\pm \nu$ with $\ell = e,\mu$, suppressing SM processes with purely hadronic final states. The other $W_h^\pm$ decays hadronically, leading to the total final state containing one lepton, missing transverse energy ($E_{T}^{\text{miss}}$) and jets. With $\mathcal{L} = 140~\text{fb}^{-1}$ of Run~2 data, ATLAS performed a search~\cite{ATLAS:2024gyc} for such a final state, placing the lower bound $m_T > 1.36$~TeV (95\%~CL). 

Instead, in the large $Z'$ coupling limit, the most stringent signature is $pp \to T\bar{T}\to t\bar{t}Z'Z'\to t\bar{t}+E_T^{\text{miss}}$. We note that the $Z'$ is expected to decay mostly invisibly in the $r_\psi = 12$ scenario, with $\mathcal{B}(Z'\to \psi\bar{\psi}) \approx 1$. The $t\bar{t}+E_{\text{miss}}$ final state is the focus of a number of ATLAS and CMS analyses, e.g., searches for pair-produced stops~\cite{ATLAS:2024rcx, ATLAS:2022ihe, ATLAS:2020dsf}. As there is no dedicated search for vector-like top partners with this final state, we use the \texttt{CheckMATE}~\cite{Drees:2013wra} package to recast these analyses.
To this end, we create a \texttt{FeynRules}~\cite{Alloul:2013bka} model file to generate a UFO output for the model with a dark fermion. We then simulate $pp\to T\bar{T}$ events at NLO in QCD using \texttt{MadGraph5\_aMC@NLO}~\cite{Alwall:2014hca} and $T$ decays using \texttt{MadSpin}~\cite{Artoisenet:2012st}, followed by showering and hadronisation with \texttt{Pythia}~8~\cite{Sjostrand:2007gs}. We rescale the obtained cross-sections to the full NNLO QCD results, as computed by \texttt{top++}~\cite{Czakon:2011xx}. First, we simulate the case of both top partners decaying to a top quark and $Z'$ (which most likely occurs on the right-hand side of Fig.~\ref{fig:BR_plot_top}) for $T$ masses in the range $m_T \in [1,2]$~TeV. We denote the obtained cross-sections as $\sigma(pp \to (T\to t Z')(\bar{T}\to \bar{t}Z'))= \sigma_{tZ'}^{}$. At each mass point we use \texttt{CheckMATE} to find the 95\%~CL upper limit on the cross section, constructing the ratio $\hat{\sigma}_{tZ'} = \sigma_{tZ'}^{\text{th}}/\sigma_{tZ'}^{95\%}$. We find that the scenario is excluded, i.e. $\hat{\sigma}_{tZ'} > 1$, for $m_T < 1.71$~TeV (95\%~CL) by the ATLAS search in Ref.~\cite{ATLAS:2020dsf}. Furthermore, we simulate the scenario of one top partner decaying to a top quark and $Z'$ and the other to SM final states ($bW^+, tZ, th$), which is most probable in the transition region of Fig.~\ref{fig:BR_plot_top}. We likewise define the cross-section $\sigma(pp \to (T\to tZ/th/bW^+)(\bar{T}\to \bar t Z')) + \sigma(pp \to (T\to tZ')(\bar{T}\to \bar tZ/\bar th/ \bar bW^-))= \sigma_{\text{SM}/tZ'}^{}$ obtain the ratio $\hat{\sigma}_{\text{SM}/tZ'}$ for this scenario from \texttt{CheckMATE}. We find $\hat \sigma_{\text{SM}/tZ'}>1$ for $m_T < 1.56$~TeV (95\%~CL) by the ATLAS search in Ref.~\cite{ATLAS:2020dsf}. Finally, for the scenario where both top partners decay to SM final states, we extract the ratio $\hat{\sigma}_{\text{SM}}$ from Fig.~5(b) of Ref.~\cite{ATLAS:2024gyc}. Now, we can exclude the parts of the parameter space at $95\%$~CL where the ratio
\begin{align}
R &= \hat{\sigma}_\text{SM}\mathcal{B}_\text{SM}^2 + \hat{\sigma}_{\text{SM}/tZ'}\mathcal{B}_\text{SM}\mathcal{B}_{tZ'} + \hat{\sigma}_{tZ'}\mathcal{B}_{tZ'}^2 \,,
\end{align}
satisfies $R > 1$, where $\mathcal{B}_{X}=\mathcal{B}(T\to X)$. The resulting exclusion is depicted as a dark blue region in Fig.~\ref{fig:gtqp_plot}. For $\tilde{g}q'$ values closer to the consistency constraint, $m_T > 1.36$~TeV, and for larger $\tilde{g}q'$ values, $m_T > 1.71$~TeV; this is expected from the scaling of $y_T^t$ in the parameter space to explain the Belle~II excess.

\subsubsection{Higgs Measurements}

Another important constraint on the model arises from measurements of the top Yukawa coupling $y_t$. In our setup, $y_t$ is determined from the input parameters $\{m_T,\tilde{g}q', y_T^t\}$ as $vy_t/\sqrt{2} = m_t c_L c_R + m_T s_L s_R$. Thus, the impact of $t-T$ mixing is a positive shift to the SM value $y_t^{\text{SM}}$, such that the quantity $\kappa_t = y_t/y_t^{\text{SM}}$ must lie in range $\kappa_t\in[1,\sqrt{m_T/m_t}]$. The strongest direct constraints on $\kappa_t$ are from CMS~\cite{CMS:2020mpn}, which searched for $pp\to t\bar{t}h$ followed by $h\to \tau^+\tau^-/W^+W^-/ZZ$, targeting final states with multiple electrons, muons or hadronically decaying tau leptons. If the tau lepton Yukawa coupling is the same as the SM value, CMS finds $\kappa_t\in [0.7, 1.1]$ at 95\%~CL. The upper limit $\kappa_t < 1.1$ excludes a significant portion of the $(m_T,\tilde{g}q')$ parameter space, shown as a light green region in Fig.~\ref{fig:gtqp_plot}. Combining Eq.~\eqref{eq:BelleII_approx} with the expression for $y_t$ in terms of the input parameters, we can derive that
\begin{align}
\label{eq:top_Yukawa_bound}
\tilde{g}q' \lesssim 7.4 \times 10^{-5}\bigg[\frac{8.6}{X_t}\bigg] \,,
\end{align}
is approximately excluded by the CMS $\kappa_t$ measurement.

Finally, we comment on constraints on the model from Higgs decays. The potentially relevant interactions of the Higgs boson in the model are
\begin{align}
\label{eq:Higgs_interactions}
\mathcal{L} &\supset - s_\phi M_{Z'}^2 \frac{h}{\tilde{v}} Z_\mu' Z'^\mu + \frac{e}{16\pi^2} c_{h\gamma Z'}\frac{h}{v} F_{\mu\nu}Z'^{\mu\nu}\,.
\end{align}
The first is induced by $h - \phi$ mixing and can contribute to the decay $h\to Z' Z'$, with $\Gamma(h\to Z'Z') = m_h^3 s_\phi^2/32\pi \tilde{v}^2$, and to invisible Higgs decays ($h\to\text{inv.}$) if $Z'$ decays are invisible. However, the scalar mixing depends on the value of $M_\phi$, which we do not fix in Fig.~\ref{fig:gtqp_plot}, but require to lie in the range in Eq.~\eqref{eq:Mphi_allowed} compatible with perturbative unitarity. For example, for $m_T = 10$~TeV and $y_T^t = 1$, the value of $\tilde{g}q'$ implied by Belle~II is just below the upper bound from BaBar. The allowed $\phi$ mass range here is $M_\phi\in[1.1,4.2]$~TeV. We find that the upper bound $\mathcal{B}(h\to \text{inv.})< 0.107$ (95\%~CL)~\cite{ATLAS:2023tkt} excludes this scenario only for $M_\phi < 1.9$~TeV. In the parameter space not excluded by the other constraints, we find that there are always values of $M_\phi$ not excluded by the constraint from $h\to\text{inv.}$. The second term in Eq.~\eqref{eq:Higgs_interactions} is generated in the model via the kinetic and mass mixing and one-loop diagrams containing $t/T$, and induces $h\to\gamma Z'$ followed by $Z'\to\text{inv.}$. However, we verify that the bound $\mathcal{B}(h\to \gamma+\text{inv.}) < 0.013$~\cite{ATLAS:2024cju} is not competitive in the parameter space of Fig.~\ref{fig:gtqp_plot}.

\subsubsection{Electroweak Precision Observables}

%
\begin{figure}[t!]
  \centering
  \includegraphics[width=0.445\columnwidth]{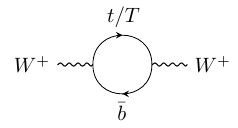}
  \includegraphics[width=0.405\columnwidth]{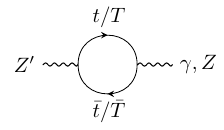}
  \caption{One-loop contributions of the top partner to the self-energies of the SM gauge fields.}
  \label{fig:STU}
\end{figure}

As shown in Fig.~\ref{fig:STU}, the heavy top partner contributes to the self-energies of the SM gauge fields and therefore to the parameters $S$, $T$ and $U$ entering the predictions for electroweak precision observables (EWPOs), given by
\begin{align}
S &= -\frac{1}{2\pi}s_L^2\bigg[\frac{1}{3}\ln r + c_L^2 f(r)\bigg] \,, \nonumber \\
T  &= -\frac{3}{8\pi s_w^2} \frac{m_t^2}{M_W^2} s_L^2 \Big[1 - r + c_L^2 g(r)\Big] \,, \nonumber \\
U  &= \frac{1}{2\pi}s_L^2\Big[\ln r + c_L^2 f(r)\Big] \,,
\label{eq:STU}
\end{align}
{where the loop functions and their arguments have been defined previously in  Sec.~\ref{subsubsec:gauge}.}
In the relevant parameter space, we find that $T\gg S,U$. We may then use the best-fit value of the $\rho$ parameter,
$\rho \approx 1+\alpha T = 1.00031 \pm 0.00019$~\cite{ParticleDataGroup:2024cfk}, which excludes the orange shaded region in Fig.~\ref{fig:gtqp_plot}. We note that the kinetic and mass mixing of $Z'$ also contribute indirectly to the EWPOs, as they induce a small shift to the SM $Z$ interactions and the input parameters $M_Z$ and $G_F$~\cite{Babu:1997st,Hook:2010tw,Curtin:2014cca}. However, this effect is expected to constrain only the parameter space with $\tilde{g}q'> 10^{-2}$, which is not considered in this work to maintain the validity of the leading order expansion used for the $Z'$ mixing. This region of parameter space is additionally excluded by the BaBar dark photon search.

Through the $Z-Z'$ mixing, the invisible $Z$ decay width can be increased from the contribution of $Z \to \psi\bar{\psi}$. The decay rate is given by Eq.~\eqref{eq:ZptoXX} with the replacements $\tilde{g}q'_\psi\to \tilde{g}q'_\psi\Delta_Z$ and $M_{Z'}\to M_Z$. The parameter space can then be constrained by the upper limit from LEP~\cite{ALEPH:2005ab}, $\Delta\Gamma_\text{inv} < 2~\text{MeV}$ (95\%~CL). However, the upper bound, which can be estimated as 
\begin{align}
\tilde{g}q' \lesssim 0.12 \bigg[\frac{12}{r_\psi}\bigg]^{1/2}\bigg[\frac{L_T}{55}\bigg]^{-1/2} \,,
\end{align}
is only stronger than the BaBar bound for very large values of the $U(1)'$ charge ratio, $r_\psi\gtrsim 10^{4}$, and therefore does not appear in Fig.~\ref{fig:gtqp_plot}. Other related probes which we have verified are less sensitive than the BaBar bound are the anomalous magnetic moment of the muon as well as atomic parity violation in Cesium~\cite{Davoudiasl:2012ag}.

\subsubsection{Other Flavour Probes}

Here we briefly review other potential flavour probes of the model and discuss why they are currently less sensitive than the bounds shown in Fig.~\ref{fig:gtqp_plot}. In principle, $Z'$ can feature as a real or virtual state in other rare processes, since it couples to FCNC down-type quark currents in Eq.~\eqref{eq:vector_coupling}. However, for $M_{Z'} = 2.1~\text{GeV}$, $Z'$ cannot be produced on-shell from kaon decays. While $K\to \pi \psi\bar{\psi}$ via an off-shell $Z'$ is kinematically possible for $2m_\psi < m_K - m_\pi$, we show in the next section that the viable region for $\psi$ is outside this range. In particular being stable and weakly coupled, $\psi$ will necessarily decouple from the primordial plasma in the early universe and form a thermal relic. In order not to overclose the universe, its annihilation cross-section should thus be sizeable enough. As we show in the next section, for $m_\psi < M_{Z'}/2$ as required by Belle II and LHCb, this is only possible close to the resonant condition $0.9 \lesssim m_\psi/\text{GeV}\lesssim 1$.
At the boundary of the allowed region, $\psi$ can actually constitute all of the observed DM. Thus, we do not consider much lower $m_\psi$ values.

The processes $B\to \pi Z'/\rho Z'$ are kinematically allowed, but suppressed by $V_{td}/V_{ts}$ with respect to $B\to K^{(*)} Z'$, and subject to similar upper bounds as $B\to K^* E_{\text{miss}}$~\cite{Belle:2017oht}. 
Finally, the heavy top partner $T$ contributes at the one-loop level to rare SM processes via the charged- and neutral-current interactions in Eq.~\eqref{eq:top_currents}. For example, the $b\to s\gamma$ transitions as well as $B_{d,s}$ meson oscillations are modified in the model. However, since the contributions of $T$ are always proportional to $s_L\ll s_R$, the current sensitivities are insufficient to constrain the relevant parameter space of the model in Fig.~\ref{fig:gtqp_plot}. In particular, EWPO are always more constraining, see Ref.~\cite{Fajfer:2013wca} for a more complete discussion.

\subsection{Dark Matter}
\label{subsec:DM}

The precise CMB measurements by the Planck collaboration constitute an important constraint on the model, due to the presence of a stable $\psi$ in the early universe. Incidently, the measured relic abundance of DM  $\Omega_{\text{DM}}h^2 = 0.120 \pm 0.001$~\cite{Planck:2018vyg}, can be readily accommodated in the model, with $\psi$ constituting (part or) all of the DM ($\Omega_\psi \leq \Omega_{\text{DM}}$). Thus the measured DM relic abundance can be considered both as an opportunity and a constraint on the model.

The freeze-out of $\psi$ in the early universe is governed by the Boltzmann equation,
\begin{align}
\label{eq:Boltzmann}
\frac{dY_\psi}{d x} = - \frac{s \braket{\sigma v}}{H x Z}\Big(Y_\psi^2 - Y_{\psi,\text{eq}}^2 \Big) \,,
\end{align}
where $x = m_\psi/T$, with $T$ being the temperature of the thermal bath. Furthermore, $Y_\psi = n_\psi/s$ is the number density of $\psi$ in units of the entropy density $s = 2\pi^2h_{\text{eff}}T^3/45$ (with $Y_{\psi,\text{eq}}$ the equilibrium value), $H = \sqrt{8\pi \rho/3M_{\text{Pl}}^2}$ is the Hubble expansion rate with a radiation-dominated energy density $\rho = \pi^2 g_{\text{eff}} T^4/30$,  $\braket{\sigma v}$ is the thermally averaged annihilation cross section of $\psi$ into species still in thermal equilibrium, and $Z = (1 + \frac{1}{3}d\ln h_{\text{eff}}/d\ln T)^{-1}$ accounts for the rate of change of $T$. Above, $M_{\text{Pl}} = 1.22\times 10^{19}$~GeV is the Planck mass and $g_{\text{eff}}$ ($h_{\text{eff}}$) is the effective number of degrees of freedom for the energy (entropy) density.

We may solve Eq.~\eqref{eq:Boltzmann} approximately as
\begin{align}
\frac{1}{Y_\psi(x)} - \frac{1}{Y_f} = \sqrt{\frac{\pi}{45}}M_{\text{Pl}} m_\psi\int_{x_f}^x dx\frac{\sqrt{g_*}\braket{\sigma v}}{x^2} \,,
\end{align}
where $Y_f \equiv Y_{\psi,\text{eq}}(x_f)$, with $x_f$ the freeze-out value of $x$, and $g_* \equiv h_{\text{eff}}^2/g_{\text{eff}}Z^2$. Then, the present day relic density of $\psi$ can be determined as $\Omega_\psi h^2 = m_\psi s_0 Y_0/\rho_{\text{cr},0}$, with $Y_0 \equiv Y_\psi(x_0)$, or
\begin{align}
\label{eq:Omega_approx}
\frac{1}{\Omega_{\psi}h^2} = \int_{x_f}^x dx\frac{\sqrt{g_*}}{x^2}\frac{\braket{\sigma v}}{2.0\times 10^{-27}~\text{cm}^3~\text{s}^{-1}}\,,
\end{align}
where the critical energy density is $\rho_{\text{cr}} = 3M_{\text{Pl}}^2H^2/8\pi$ and we account for non-identical DM particles ($\psi$) and antiparticles ($\bar{\psi}$).

\begin{figure}[t!]
  \centering
  \includegraphics[width=0.32\columnwidth]{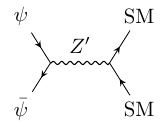}
  \includegraphics[width=0.32\columnwidth]{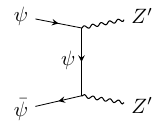}
  \caption{Scattering processes contributing to the freeze-out of the DM candidate $\psi$.}
  \label{fig:DM_annihilation}
\end{figure}

We are interested in the freeze-out of $\psi$ for masses which can accommodate the Belle~II excess, i.e. $m_\psi < M_{Z'}/2$. The dominant annihilation process in this mass range is $\psi\bar{\psi}\to \text{SM} + \text{SM}$ through $s$-channel $Z'$ exchange, shown in Fig.~\ref{fig:DM_annihilation} (left). The $Z'$ couples at tree-level to $\psi$ and via the kinetic and mass mixing to light SM fields. As the annihilation can take place near the narrow $Z'$ resonance, the thermally averaged cross section cannot be estimated using the typical non-relativistic expansion in the relative DM velocity $v$. The thermally averaged Breit-Wigner cross section is s-wave dominated and given at leading order in $\epsilon_\psi \equiv (M_{Z'}^2 - 4m_\psi^2)/4m_\psi^2$ by~\cite{Gondolo:1990dk} 
\begin{align}
\label{eq:sv_resonant}
\braket{\sigma v} = \frac{12\pi^{3/2}\gamma_{Z'}\mathcal{B}_{Z'}}{m_\psi^2} x^{3/2}e^{-x\epsilon_\psi}\,,
\end{align}
defining $\gamma_{Z'} \equiv M_{Z'}
\Gamma_{Z'}/4m_{\psi}^2$ and $\mathcal{B}_{Z'} \equiv \mathcal{B}_{\psi\bar{\psi}}\mathcal{B_{\text{SM}}}$, where $\mathcal{B}_{\psi\bar{\psi}} = \Gamma_{\psi\bar{\psi}}/\Gamma_{Z'}$. With the vector-like fermion $\psi$ as the only light dark sector particle, $\mathcal{B}_{\text{SM}} = 1 - \mathcal{B}_{\psi\bar{\psi}}$. Inserting Eq.~\eqref{eq:sv_resonant} into Eq.~\eqref{eq:Omega_approx} yields the approximate analytical result
\begin{align}
\label{eq:Omega_approx_2}
\Omega_{\psi}h^2 = \frac{3.4\times 10^{-12}~\text{GeV}}{\Gamma_{Z'}} \frac{\sqrt{\hat{\epsilon}_\psi}}{\mathcal{B}_{Z'}}\,,
\end{align}
with $\hat{\epsilon}_\psi \equiv \epsilon_\psi/\bar{g}_*(1+\epsilon_\psi)^4\text{erfc}^2(\sqrt{x_f \epsilon_\psi})$, where $\text{erfc}$ is the complementary error function, $g_*$ is fixed to its value at chemical decoupling, $\bar{g}_* = g_*(x_f)$, and we always assume that $M_{Z'} = 2.1$~GeV. In the limit where $\mathcal{B}_{\psi\bar{\psi}} \approx 1$ and the kinetic mixing $\epsilon_A$ dominates the $Z'$ coupling to SM states, which generally holds when the Belle~II excess is satisfied, Eq.~\eqref{eq:Omega_approx_2} gives
\begin{align}
\label{eq:Omega_approx_3}
\frac{\Omega_\psi}{\Omega_{\text{DM}}} = \bigg[\frac{1.5\times 10^{-4}}{\tilde{g}q'}\bigg]^2\bigg[\frac{55}{L_T}\bigg]^{2}\bigg[\frac{\hat{\epsilon}_\psi}{2.3}\bigg]^{1/2}\,,
\end{align}
using $x_f = 20$. Note that the result is independent of the $U(1)'$ charge ratio $r_\psi$. Using Eq.~\eqref{eq:BelleII_approx}, the Belle~II excess and the correct DM abundance can both be recovered for a specific value of $m_T$ when the top partner Yukawa is
\begin{align}
y_T^t = 0.29\bigg[\frac{m_T}{10~\text{TeV}}\bigg] \bigg[\frac{55}{L_T}\bigg]^{1/2}\bigg[\frac{8.6}{X_t}\bigg]^{1/2}\bigg[\frac{\hat{\epsilon}_\psi}{2.3}\bigg]^{1/8}\,.
\end{align}
To verify these estimates, we use the \texttt{micrOMEGAs}~\cite{Alguero:2023zol} package to compute $\Omega_\psi$ by numerically solving Eq.~\eqref{eq:Boltzmann}. We use \texttt{FeynRules}~\cite{Alloul:2013bka} to write \texttt{CalcHEP}~\cite{Belyaev:2012qa} files for the same model file used in the collider analysis in Sec.~\ref{subsec:pheno_DM}, which are then implemented in \texttt{micrOMEGAs}.

\begin{figure}[t!]
  \centering
  \includegraphics[width=\columnwidth]{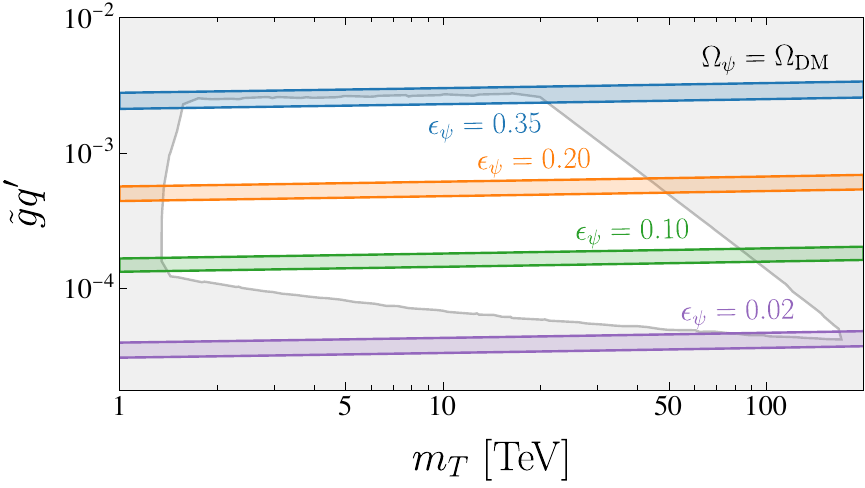}
  \includegraphics[width=\columnwidth]{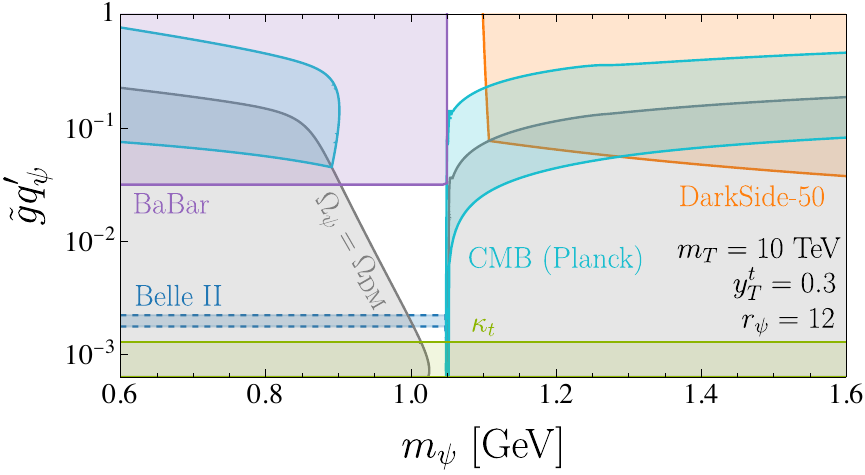}
  \caption{(Above) The $(m_T, \tilde{g}q')$ parameter space where the correct DM relic abundance ($\Omega_\psi = \Omega_\text{DM}\pm 20\%$) is obtained for different values of $\epsilon_\psi$ (color shaded bands). The unshaded region is the parameter space not excluded in Fig.~\ref{fig:gtqp_plot} which can accommodate the Belle~II excess. (Below) Bounds on the light $\psi$ DM scenario in the $(m_\psi, \tilde{g}q_\psi')$ plane. The correct DM relic abundance is found along the gray line, with DM overabundance obtained below. For $m_\psi < M_{Z'}/2$, there is a window where Belle~II and Planck can be satisfied and the direct and indirect DM bounds are evaded. We show the constraints from BaBar (purple), the top Yukawa coupling (light green), Planck (light blue) and DarkSide-50 (orange).}
  \label{fig:DM_plot}
\end{figure}

In Fig.~\ref{fig:DM_plot} (above), we show where in the $(m_T,\tilde{g}q')$ plane the correct DM relic abundance is obtained for four benchmark $\epsilon_\psi$ values, with $\epsilon_\psi = \{0.35, 0.20,0.10,0.02\}$ corresponding to $m_\psi \approx \{0.90,0.96,1.00,1.04\}$~GeV. The bands indicate $\Omega_\psi = \Omega_{\text{DM}}\pm 20\%$. Firstly, we find that $\Omega_\psi$ is well approximated by the scaling in Eq.~\eqref{eq:Omega_approx_3} in this regime. Comparing the DM predictions to the region still allowed by the current constraints in Fig.~\ref{fig:gtqp_plot}, which are combined as a gray shaded region in Fig.~\ref{fig:DM_plot} and are approximately independent of $\epsilon_\psi$, lower and upper limits on $\epsilon_\psi$ are implied. A lower bound on $\epsilon_\psi$ is determined by the CMS bound on the top Yukawa coupling. Combining Eqs.~\eqref{eq:top_Yukawa_bound} and \eqref{eq:Omega_approx_3}, we find
\begin{align}
\hat{\epsilon}_\psi \gtrsim 0.13 \bigg[\frac{L_T}{55}\bigg]^4 \bigg[\frac{8.6}{X_t}\bigg]^4 \,,
\end{align}
which gives $\epsilon_\psi \gtrsim \{0.05, 0.02\}$ for $m_T = \{10,150\}$~TeV, using the definition of $\hat{\epsilon}_\psi$ below Eq.~\eqref{eq:Omega_approx_2}. Upper bounds on $\epsilon_\psi$ are determined either by the BaBar bound from $e^+e^- \to \gamma Z'$ or the perturbative unitarity bound on the scalar sector. From the former, combining Eq.~\eqref{eq:BaBar_bound} with Eq.~\eqref{eq:Omega_approx_3} yields $\epsilon_\psi \lesssim 0.35$, independent of $m_T$. The latter instead implies, using Eq.~\eqref{eq:pert_bound},
\begin{align}
\hat{\epsilon}_\psi \lesssim 2.7\times 10^{7}\bigg[\frac{10~\text{TeV}}{m_T}\bigg]^8\bigg[\frac{X_t}{8.6}\bigg]^4\,,
\end{align}
giving $\epsilon_\psi \lesssim \{0.47,0.04\}$ for $m_T = \{10,150\}$~TeV. Thus, we find that for viable DM, $0.02 \lesssim \epsilon_\psi \lesssim 0.35$ is required.

Before moving on, we note that in the regime $\epsilon_{\psi} \ll 1$, i.e., very close to the $Z'$ resonance, the abundance of $\psi$ evolves long after chemical decoupling at $x_f$ in a belated freeze-out process~\cite{Belanger:2025kce}. Kinetic equilibrium between $\psi$ and the SM bath cannot be maintained during this freeze-out, because non-resonant DM-SM scattering processes are suppressed. Thus, solving the Boltzmann equation in Eq.~\eqref{eq:Boltzmann} no longer provides reliable results for $\Omega_\psi h^2$~\cite{Binder:2017rgn}. The impact of kinetic decoupling on the DM abundance can be estimated, following the approach Ref.~\cite{Belanger:2025kce}, by multiplying Eq.~\eqref{eq:Omega_approx_2} by the factor $k_{\text{dec}} = 2\sqrt{\pi x_d\epsilon_\psi}$, where $x_d = m_\psi/T_d$ with $T_d$ the temperature of kinetic decoupling. For $\gamma_{Z'}\lesssim 10^{-7}$, which is anticipated in the model for $\epsilon_\psi\ll 1$, $x_d \approx x_f$. However, we still find that $k_{\text{dec}} > 1$ for the $\epsilon_\psi$ values in the range $0.02 \lesssim \epsilon_\psi \lesssim 0.35$, so the DM abundance is not suppressed by the kinetic decoupling.

To illustrate further the viable DM parameter space, we show in Fig.~\ref{fig:DM_plot} (below) the relevant constraints in the $(m_\psi, \tilde{g}q_\psi')$ plane for the benchmark values $m_T = 10$~TeV, $y_T^t = 0.3$ and $r_\psi = 12$. Unlike Figs.~\ref{fig:gtqp_plot} and \ref{fig:DM_plot} (above), we do not require the Belle~II excess to be accommodated at each point in the parameter space. For $m_\psi < M_{Z'}/2$, the favoured region for Belle~II is shown as a blue band. We also extend the mass range to $m_\psi > M_{Z'}/2$, where additional invisible $Z'$ decays would need to be present to explain the Belle~II result and be compatible with LHCb. We restrict to $m_\psi < M_{Z'}$ so that the $s$-channel process $\psi\bar{\psi}\to \text{SM} + \text{SM}$ still dominates the DM freeze-out. Firstly, the correct DM relic abundance is obtained along the gray solid line, while the gray shaded region below indicates the overproduction of DM. We show the bounds from BaBar and $\kappa_t$ discussed in Secs.~\ref{subsubsec:darkPhoton} and~\ref{subsubsec:collider} which constrain $\tilde{g}q'$, in addition to the limits from (in)direct detection discussed below.

Relevant direct detection bounds in this mass range come from DarkSide-50, constraining DM ($\psi$) scattering on Ar~\cite{DarkSide:2018bpj}. For the parameter space shown in Fig.~\ref{fig:DM_plot}, the kinetic mixing $\epsilon_A$ dominates the interactions of $Z'$ with the SM states. Thus, spin-independent (SI) scattering of $\psi$ and protons via the exchange of $Z'$ dominates the DM-Ar scattering cross section. Using \texttt{micrOMEGAs}, we compute at each point in the parameter space the ratio of the expected SI cross section to the 90\%~CL excluded value, $\hat{\sigma}_{\text{SI}} = \sigma_{\text{SI}}^{\text{th}}/\sigma_{\text{SI}}^{90\%}$, excluding $\hat{\sigma}_{\text{SI}} > 1$ at 90\%~CL. If $\psi$ makes up a faction of the observed DM abundance, we exclude $\hat{\sigma}_{\text{SI}}(\Omega_\psi/\Omega_{\text{DM}}) > 1$. The orange shaded region in Fig.~\ref{fig:DM_plot} shows the resulting limit, excluding viable DM scenarios with $m_\psi > 1.1$~GeV in the mass range shown. Importantly, the most relevant region $m_\psi < M_{Z'}/2 $ remains unconstrained.\footnote{There are also direct detection bounds from DM ($\psi$) scattering on electrons via $Z'$ exchange, with XENON1T~\cite{XENON:2019gfn} and PandaX-4T~\cite{PandaX:2022xqx} being the most stringent experiments. Similar to the nuclear scattering, the bounds are not relevant in the resonant regime in Fig.~\ref{fig:DM_plot}.}

At recombination, the annihilation of $\psi$ into SM states provides an indirect detection bound on the model. CMB anisotropies measured by Planck constrain the parameter $p_{\text{ann}} < 3.2 \times 10^{-28}~\text{cm}^3~\text{s}^{-1}~\text{GeV}^{-1}$~(95\%~CL)~\cite{Planck:2018vyg}, which is sensitive to energy injection from $e^\pm$ and $\gamma$ produced by DM annihilation final states. We can estimate 
\begin{align}
p_{\text{ann}} = \sum_i\frac{f_i^{\text{eff}} \braket{\sigma v}_i}{m_\psi}\bigg(\frac{\Omega_\psi}{\Omega_{\text{DM}}}\bigg)^2 \,,
\end{align}
where, for the annihilation channel $i$, $f_i^{\text{eff}}$ is the fraction of energy transferred to the plasma~\cite{Slatyer:2015jla,Leane:2018kjk}, $\braket{\sigma v}_i$ is the thermally averaged cross section, and we re-scale by the fractional contribution of $\psi$ to the total DM relic density. The scattering cross section is well approximated by the zero velocity limit expression,
\begin{align}
\braket{\sigma v}_i = \frac{6\pi \gamma_{Z'}^2\mathcal{B}_{\psi\bar{\psi}}\mathcal{B}_i}{m_\psi^2\sqrt{\epsilon_\psi}(\gamma_{Z'}^2 + \epsilon_{\psi}^2)} \,.
\end{align}
Using \texttt{micrOMEGAs}, the $f_i^{\text{eff}}$ and $\braket{\sigma v}_i$ are computed and the parameter space excluded by the Planck upper limit is shown as a light blue shaded region in Fig.~\ref{fig:DM_plot}. For $\epsilon_\psi$ sufficiently large, and thus below the resonance, the $\braket{\sigma v}_i$ are suppressed enough for the CMB bound to be evaded. Note that for masses in this range, the $e^+$ and $\gamma$ spectra calculated in \texttt{micrOMEGAs}, which act as input to $f_i^{\text{eff}}$, carry significant uncertainties. Hadronisation of the quark final-states is calculated using $\texttt{Pythia}$ and extrapolated to lower masses, where hadronic resonances make the $\texttt{Pythia}$ calculations less reliable. Nevertheless, we do not expect this to make up for the large suppression of $\braket{\sigma v}_i$ below the resonance at recombination. {The late time annihilation of DM to SM particles is similarly suppressed, and therefore indirect detection bounds, e.g. from Super-Kamiokande~\cite{Super-Kamiokande:2020sgt} or $X$-ray data~\cite{Cirelli:2023tnx,Balaji:2025afr}, are evaded.}

To conclude, the resonant thermal annihilation of $\psi$ in the early universe for $0.9 \lesssim m_\psi/\text{GeV}\lesssim 1.0~\text{GeV}$, where $\psi$ simultaneously appear as missing energy at Belle~II, offers a compelling DM scenario which is not yet experimentally excluded by direct or indirect detection. The model presented offers the minimal field content to fit both the Belle~II data and generate the correct relic abundance with a light $Z'$. Of course, the presence of additional fields in an extended dark sector charged under $U(1)'$ would modify the DM phenomenology, as we discuss briefly in the next section.

%
\section{Discussion and Conclusions}
\label{sec:conclusions}

In previous work~\cite{Bolton:2024egx,Bolton:2025fsq}, we demonstrated within a
model–independent EFT framework that the Belle~II
excess in $B^+ \to K^+ E_{\rm miss}$ can be well described by the
emission of a single light vector boson, with a mass of about
$M_{Z'} \simeq 2.1$~GeV, in $b\to s$ transitions.
We have now extended this to a
simple UV-complete construction, based on a
Higgsed $U(1)'$ gauge symmetry with a light vector boson $Z'$, a complex scalar $\Phi$ whose vacuum
expectation value gives mass to $Z'$ and a
$U(1)'$ charged vector-like top partner $T$ realising a rank-1 pattern of flavour breaking, aligned to the third generation, that naturally enhances $b\to s$ processes.

In this minimal construction, loop-induced $b\to s Z'$ transitions generate the desired
Belle~II signal {through a left handed current operator}, but the same loops unavoidably induce kinetic and mass
mixing between $Z'$ and the electroweak gauge bosons, leading to
irreducible couplings of $Z'$ to charged leptons that cannot even be fine-tuned away. 
The resulting branching ratios  in the minimal model,
displayed in Fig.~\ref{fig:BR_plot_Zprime} , show that $Z'$ decays dominantly into
visible SM final states, with a sizeable $\mu^+\mu^-$
branching fraction in the parameter region favoured by Belle~II.
The corresponding contributions to $B\to K^{(*)}\mu^+\mu^-$ are excluded
by existing LHCb searches for prompt and displaced dimuon resonances
(Fig.~\ref{fig:LHCb}).
We therefore conclude that the minimal aligned $U(1)'$ model cannot
account for the Belle~II anomaly; its failure is directly tied to the
absence of sufficiently large invisible $Z'$ decay channels {as the $Z'$ effectively always decays inside the detector}.

Motivated by this, we extended the $U(1)'$ charged sector by adding a light
SM singlet fermion $\psi$, which opens a dominant invisible decay mode $Z'\to\psi\bar{\psi}$ and sufficiently suppresses the
visible branching ratios (provided the $\psi$ $U(1)'$ charge is an order of magnitude larger than that of $T$ and $\Phi$), as seen in the modified decay patterns in
Fig.~\ref{fig:BR_plot_Zprime-2}.
The same interaction controls the $\psi$ annihilation rate in the early universe, and a
combination of flavour, collider, electroweak, and cosmological constraints
selects a narrow $\psi$ mass regime, just below half
of the $Z'$ mass, illustrated in
Fig.~\ref{fig:DM_plot}, where $\psi$ can constitute part or all of DM.
Outside this resonant region, the model violates at least one of these constraints. Future improved measurements and searches, planned at existing and upcoming experimental facilities like 
Belle II~\cite{Belle-II:2018jsg} and FCC~\cite{FCC:2018byv}, will be able to further constrain the viable parameter space. The model thus represents a highly constrained and predictive scenario relating missing energy signatures in rare $B$ decays to a comprehensive collider and DM phenomenology. 

Extensions of the model field content could potentially relax some of the constraints found in the minimal model. For instance, the large $U(1)'$ charge ratio ($r_\psi$) between $\psi$ and $T,\Phi$  could be avoided by  considering multiple light dark fermions $\psi_i$ with $r_{\psi_i}\sim 1$ instead of one $\psi$ with $r_\psi\gg 1$. Secondly, the CMB and direct detection constraints could be further relaxed for dark fermions with axial-vector couplings to the $Z'$. Such a scenario would be obtained if different $U(1)'$ charges are assigned to the left- and right-handed components $\psi_L \equiv \psi$ and $\psi_R \equiv \psi'$. For example, with $q'_L \equiv q'_\psi \neq q'_R=q'_{\psi'}$, we have
\begin{align}
\label{eq:p-wave_model}
\mathcal{L}\supset \bar{\psi} i\slashed{D} \psi + \bar{\psi'} i\slashed{D} \psi' - \frac{1}{2}\Big[Y_\psi \bar{\psi'}^c \psi' \tilde{\Phi} + \text{h.c.}\Big] \,,
\end{align}
where $\psi^c = C\bar{\psi}^T$, with $C$ the charge conjugation matrix, and we also introduce the scalar $\tilde{\Phi} = (\tilde{v}_\psi + \varphi + ia)/\sqrt{2}$ with the $U(1)'$ charge $-2 q'_R$. In the broken $U(1)'$ phase, the Majorana mass $m_{\psi'} = \tilde{v}_\psi Y_\psi/\sqrt{2}$ is generated for $\psi'$. The state $\psi$ may also be a Majorana fermion if the charges obey $q' = q'_L - q'_R$, thus permitting the Yukawa coupling
\begin{align}
\mathcal{L} \supset - y_\psi \bar{\psi}\psi' \Phi + \text{h.c.}\,.
\end{align}
For $\tilde{v}y_\psi\ll \tilde{v}_\psi Y_\psi$, a seesaw-like mechanism generates a Majorana mass for $\psi$ of size $m_\psi \approx (\tilde{v}y_\psi)^2/(\sqrt{2}\tilde{v}_\psi Y_\psi)$.
In this setup, both $\psi$ and $\psi'$ have purely axial-vector couplings to $Z'$ and
their annihilation would be p-wave ($v^2$-suppressed), reducing $\braket{\sigma v}$ at recombination by $\sim(v_{\rm CMB}/v_{\text{fo}})^2 \sim 10^{-6}$, while maintaining efficient freeze-out. The annihilation then proceeds either through the $Z'$ resonance for $m_{\psi^{(\prime)}}\lesssim M_{Z'}/2$ or alternatively for $m_{\psi^{(\prime)}}\gtrsim M_{Z'}$ through the process $\psi^{(\prime)} \psi^{(\prime)} \to Z'Z'$ shown in Fig.~\ref{fig:DM_annihilation} (right). In such an extended scenario, the missing energy at Belle~II can either be provided by the $\psi^{(\prime)}$ fields, $Z'\to \psi^{(\prime)} {\psi}^{(\prime)}$, or alternatively the physical CP-even and -odd components of $\Phi$ and $\tilde{\Phi}$ (mass eigenstates $\varphi'$ and $a'$), $Z'\to \varphi' a'$, if they are sufficiently light. We leave such considerations for future work.

%
\begin{acknowledgments}
The authors would like to thank Svjetlana Fajfer for her contributions in the initial stages of this work, and Zara Barbari\'c for her assistance in deriving the collider bounds on the model. We thank Alexander Belyaev for his assistance with MicrOmegas in the early stages of the project. We also thank Genevieve Belanger and Cedric Delaunay for their insightful correspondence regarding Ref.~\cite{Belanger:2025kce} and MicrOmegas, and Marco Ardu for his comments regarding p-wave annihilation.
PDB and JFK acknowledge financial support from the Slovenian Research Agency (research core funding No. P1-0035 and N1-0321). MN acknowledges the financial support of the Spanish Government (Agencia Estatal de Investigaci\'on MCIN/AEI/10.13039/501100011033)  and the European Union NextGenerationEU/PRTR through the “Juan de la Cierva” programme (Grant No. JDC2022-048787-I) and through Grants No. PID2020-114473GB-I00 and No. PID2023-146220NB-I00. 
\end{acknowledgments}
%

\newpage
\appendix

%
\section{Full Formulae}
\label{app:exps}
%

Here, we give the full matching relations for the WET coefficients in Eq.~\eqref{eq:eff_Hamiltonian}. Firstly, the coupling of $Z'$ to the left-handed vector down-type quark current is 
\begin{align}
\label{eq:CV_full}
C_{ij}^V(M_W) &= \frac{1}{2} \mathcal{C}_{ij} m_t^2 c_L^2 \bigg[\tilde{g}q' s_R^2 C_{Z'}^V - \frac{g}{c_w} \Delta_{Z'} C_Z^V\bigg] \,,
\end{align}
at $\mu = M_W$, with
\begin{align}
C_{Z'}^V &= X(x_t) + X(x_T) + X'(x_t, x_T) + \frac{2s_L^2}{s_R^2} g(r) \,, \nonumber \\
C_Z^V &= Y(x_t) + t_L^2 r Y(x_T) - s_L^2 g(r) \,,
\label{eq:CZV}
\end{align}
where $x_{t,T} = m_{t,T}^2/M_W^2$, $r = m_T^2/m_t^2$, { the loop function $g(x)$ is defined below Eq.~\eqref{eq:delta_Z}} and we define the { additional} loop functions
\begin{align}
X(x) &= - \frac{x - 4}{x - 1} - \frac{x^2 - 2x + 4}{(x - 1)^2}\ln x\,,  \nonumber \\
X'(x, y) &= \frac{x - 4}{x - 1}\frac{2x}{x - y}\ln x + (x \leftrightarrow y)\,, \nonumber \\
Y(x) &= \frac{x - 6}{x-1} + \frac{3x + 2}{(x - 1)^2}\ln x\,. 
\end{align}
In the limit where $m_t \ll m_T$ and $\tilde{g}q'$ is sufficiently above the consistency bound in Eq.~\eqref{eq:consistency}, $s_L^2 \approx s_R^2/r\ll 1$ yields the estimates $C_{Z'}^V \approx X(x_t) + \ln x_T + 1 \approx \ln x_T - 1$, where we used $X(x_t) = -2.05$, and $C_Z^V \approx Y(x_t) = 1.47$. For $m_t \ll m_T$ and $\tilde{g}q'$ close to the consistency bound, we instead have $s_L \approx 1/\sqrt{r}$ and $s_R\approx 1$, giving $C_{Z'}^V \approx 2\sqrt{r}$ and $C_Z^V \approx Y(x_t) + 1 = 2.47$. In these two limits, $C_{ij}^V$ is given by Eqs.~\eqref{eq:CV} and \eqref{eq:CV_limit}, respectively, which are used to derive approximate results in the text. However, we use the full expression in Eq.~\eqref{eq:CV_full} for the numerical studies of this analysis.

The coupling of $Z'$ to the dipole-like down-type quark current is given at $\mu = M_W$ by
\begin{align}
\hspace{-0.8em}\,C_{ij}^T(M_W) &= \mathcal{C}_{ij} m_i  x_t c_L^2\nonumber\\
&\hspace{1.3em}\times \bigg[\tilde{g}q's_R^2 C_{Z'}^{T}- e\epsilon_A C_\gamma^T - \frac{g}{c_w}\Delta_{Z'}C_Z^T\bigg] \,,
\end{align}
with
\begin{align}
\hspace{-1em}\,C_{Z'}^T &= A(x_t) + A(x_T) + A'(x_t,x_T) + \frac{s_L^2}{s_R^2 x_t} A''(x_t,x_T) \,,\nonumber \\
\hspace{-1em}\,C_\gamma^T &= B(x_t) + t_L^2 r B(x_T)\,, \nonumber\\
\hspace{-1em}\,C_Z^T &= C(x_t) + t_L^2 r C(x_T) + \frac{s_L^2}{x_t} C''(x_t, x_T) - s_w^2 C_\gamma^T \,, 
\end{align}
where the relevant loop functions are
\begin{align}
A(x) &= \frac{x^2 - 5x - 2}{24(x - 1)^3} + \frac{x}{4(x - 1)^4}\ln x \,, \nonumber \\ 
B(x) &= \frac{8x^2 + 5x - 7}{24(x - 1)^3} - \frac{3x^2 - 2x}{4(x - 1)^4}\ln x \,, \nonumber \\ 
C(x) &= \frac{5x^2 + x}{16(x - 1)^3} - \frac{5x^2 - 3x + 1}{8(x - 1)^4}\ln x\,,
\end{align}
and
\begin{align}
A'(x,y) &= \frac{x y - 1}{12(x - 1)^2(y - 1)^2}  \nonumber \\
&\hspace{1.3em} - \frac{x^3 - 3x^2}{12(x - 1)^3(x - y)}\ln x + (x \leftrightarrow y) \,, \nonumber\\
A''(x,y) &= -\frac{5x^2 + 5x - 4}{12(x - 1)^3} + \frac{2x^2 - x}{2(x - 1)^4}\ln x \nonumber \\
&\hspace{1.3em} + \frac{4xy - 3x - 3y + 2}{6(x - 1)^2(y - 1)^2}  \nonumber \\
&\hspace{1.3em} + \frac{5x^3 - 3x^2}{6(x - 1)^3 (x - y)}\ln x + (x \leftrightarrow y) \,, \nonumber\\
C''(x,y) &= \frac{7x^2 + 10x - 5}{48(x - 1)^3} - \frac{3x^2 - x}{8(x - 1)^4}\ln x \nonumber \\
&\hspace{1.3em} - \frac{13x - 5}{48(x - 1)^2(y - 1)}  \nonumber \\
&\hspace{1.3em}  - \frac{7x^3 - 3x^2}{24(x - 1)^3 (x - y)}\ln x + (x \leftrightarrow y) \,.
\end{align}

\bibliography{main}

\end{document}